\documentclass[]{jfm}
\usepackage{graphicx}
\usepackage{newtxtext}
\usepackage{newtxmath}
\usepackage{natbib}
\usepackage{hyperref}
\usepackage{float}
\usepackage{subfig}
\usepackage{dcolumn}
\usepackage{bm}
\usepackage{gensymb}
\usepackage{tabularx}
\usepackage{color}
\usepackage{braket}
\usepackage{comment} 
\hypersetup{colorlinks=true, urlcolor=blue, citecolor=blue}

\title{Local and global bifurcations to large-scale oblique patterns in inclined layer convection}
\shorttitle{Local and global bifurcations to large-scale oblique patterns in ILC}
\shortauthor{Z. Zheng, S. Azimi, F. Reetz and T.M. Schneider}
\author{Zheng Zheng\aff{1}, Sajjad Azimi\aff{1,2}, Florian Reetz\aff{1} \and Tobias M. Schneider\aff{1}\corresp{\email{tobias.schneider@epfl.ch}}}
\affiliation{\aff{1} Emergent Complexity in Physical Systems Laboratory (ECPS), \'Ecole Polytechnique F\'ed\'erale de Lausanne, CH 1015 Lausanne, Switzerland
\aff{2} Department of Environmental Science and Engineering, California Institute of Technology, Pasadena, CA, USA}

\begin{document}
\maketitle

\begin{abstract}
In the inclined layer convection system, thermal convection in a Rayleigh--B\'enard cell tilted against gravity, the flow is subject to competing buoyancy and shear forces. For varying inclination angle ($\gamma$) and Rayleigh number ($Ra$), a variety of spatio-temporal patterns is observed. We investigate the switching diamond panes (SDP) pattern, observed at $(\gamma, Ra)\simeq(100\degree,10000)$, which exhibits large-scale oblique features and is one of the five complex tertiary patterns at Prandtl number $Pr=1.07$. First, we study the linear instability of the secondary-state transverse convection rolls and the five branches including two travelling waves and three periodic orbits, bifurcating simultaneously from it. These non-generic bifurcations arise from the breaking of the spatial $D_4$ and $O(2)$ symmetries of transverse rolls, and the resulting bifurcated solutions show large-scale diamond-shaped amplitude modulations. Second, we explore a periodic orbit that captures both the large-scale structure and small-scale defects of modulated rolls. Parametric continuation in $Ra$ reveals the global homoclinic bifurcation via which this periodic orbit emerges. Third, the boundary of the basins of attraction of two dynamically relevant periodic orbits has been characterised. Specifically, additional steady and time-periodic solutions are identified on the basin boundary and their bifurcation structures are analysed. Together, using non-linear invariant solutions and their bifurcations, we take a further step toward understanding the emergence and dynamics of SDP far from the onset of convection, where linear methods have not been applied successfully.
\end{abstract}


\section{Introduction}
\par In thermally driven inclined layer convection (ILC), a fluid layer is confined between two parallel walls inclined against gravity, and is heated from one wall while cooled from the other. The inclination angle ($\gamma$), Rayleigh number ($Ra$), and Prandtl number ($Pr$) constitute the three control parameters of the ILC system. The inclined convection system is a broadly studied model motivated by its relevance to physical processes across a wide range of applications, including geophysical, meteorological, and atmospheric flows as well as engineering systems \citep{Hart1971a, Kaushika2003, Arici2015}. In addition, this spatially extended fluid system also provides a well-controlled setting to investigate pattern formation driven by both buoyancy and shear above the onset of convection.

\par Early experimental investigations of ILC have primarily focused on heat transfer mechanisms and how the occurrence of flow instabilities dictates variations in heat transfer \citep{DeGraaf1953, Hollands1973, Ruth1980, Ruth1980a}. More recently, \citet{Daniels2000} observed and characterised a profusion of complex, self-organized, spatio-temporal patterns in a large-aspect-ratio ILC experiment at $Pr=1.07$ (compressed CO$_2$). In the ILC system, with increasing Rayleigh number, the laminar base flow loses linear stability and straight convection rolls (secondary states) form. Two types of straight rolls can emerge: longitudinal rolls resulting from buoyancy-driven primary instability and transverse rolls resulting from shear-driven instability. While these two secondary states are equilibria, the tertiary states are chaotic. \citet{Subramanian2016} studied systematically these tertiary patterns in periodic domains using Floquet analysis; specifically, they identified five tertiary patterns that show complex dynamics: skewed varicose, longitudinal subharmonic oscillations, wavy rolls, knots, and switching diamond panes (SDP, also called transverse oscillatory rolls). They reported that these five patterns can be associated to the unstable eigenmodes of the secondary states. Among these five patterns, SDP requires by far the largest domain to be observed, due to the long pattern wavelength \citep{Reetz2020}.

\par The SDP pattern exhibits large-scale oblique and diamond-shaped amplitude modulations, with the spatial phase of diamonds switching in time inducing small-scale defects in the rolls. The emergence of the SDP pattern represents the onset of complex dynamics with increasing Rayleigh number at inclination angles close to $\gamma=90\degree$, where the shear in the laminar flow is the highest. Over a relatively broad range of $Ra$ and $\gamma$ where shear forces dominate, SDP emerges as a prevailing pattern. While complex oblique large-scale patterns have been extensively studied in plane Couette flow \citep{Barkley2005, Duguet2010, Duguet2013, Reetz2019a}, in plane Poiseuille flow \citep{Fukudome2012, Tuckerman2014turbulent, Xiong2015, Tao2018}, and in counter-rotating Taylor--Couette flow \citep{Meseguer2009, Dong2009}, such patterns have been investigated to a relatively limited extent in thermal convection systems.

\par The coherent features within these turbulent patterns, which recur in space and time, suggest the existence of invariant solutions to the flow equations that underlie the observed dynamics and capture the essential pattern properties. Within the dynamical systems approach to turbulence, the evolution of the flow is envisioned as a chaotic trajectory in the infinite-dimensional state space of the flow equations. This chaotic trajectory transiently visits the neighbourhoods of unstable invariant solutions embedded in the chaotic attractor of the system. Simple unstable invariant states such as (relative) periodic orbits, equilibria, and travelling waves are thus elementary building blocks to unravel chaotic dynamics of the flow \citep{Cvitanovic1991, Kawahara2001, Kawahara2012, Chandler2013, Graham2021}. Identifying an invariant solution that exhibits the spatio-temporal pattern of interest may enable us to explain the emergence of the pattern as a consequence of the chaotic trajectory visiting such simple structures in state space, and to investigate the underlying mechanisms in a simpler setting than the chaotic attractor.

\par In the context of thermal convection, \citet{Reetz2020a} and \citet{Reetz2020} discovered many invariant solutions underlying the aforementioned convection patterns, studied the temporal transitions between these invariant states, and investigated their bifurcation structures. Regarding the SDP pattern, \citet{Reetz2020} used a minimal periodic domain supporting this large-wavelength pattern, constrained the dynamics by imposing both reflection and translation symmetries, and identified one periodic orbit which they referred to as transverse oscillations. This orbit bifurcates from transverse rolls in a Hopf bifurcation, and the neutral eigenspace at the bifurcation point is two-dimensional. Without imposed symmetries, however, four complex conjugate eigenvalue pairs cross the imaginary axis simultaneously. Furthermore, this orbit only captures part of the complex SDP pattern dynamics: it does not contain any defects in rolls which are observed in both experiments and simulations. The understanding of SDP and of the linear instability of transverse rolls remain incomplete, and the bifurcation-theoretic origin of the pattern is still unclear. These open questions have motivated the present work.

\par Pattern formation, a classical topic in non-equilibrium physics and fluid mechanics \citep{Cross1993pattern}, traces its origins to the seminal work of \citet{Turing1952} in the mid-20th century. For low-dimensional model systems, theoretical analyses have usually been carried out in the linear or weakly non-linear regimes. In high-dimensional fluid dynamical systems, the focus is often in the fully non-linear regime. Nevertheless, close to a local bifurcation point, where the dimension of the stable and unstable manifolds of the parent state changes, the dynamics of the high-dimensional system can be accurately captured by an appropriate normal form. Such normal forms are the smallest systems in terms of both the number of variables and polynomial order that are able to reproduce the behaviour of the original system near the bifurcation, and are constrained by the symmetries of the parent branch. This is  an application of equivariant bifurcation theory---theory of bifurcations in dynamical systems with symmetries---via which possible bifurcation scenarios can be predicted \citep{golubitsky1985singularities, golubitsky1988singularities, Crawford1991, chossat2000methods}. It is in this spirit and with this methodology that we will analyse the complex SDP pattern and interpret various bifurcating branches.

\par In this work, we explore the state-space structures underlying the observed SDP dynamics by computing invariant solutions to the three-dimensional Oberbeck--Boussinesq equations. We characterise the linear stability of these solutions, which include both equilibria and periodic orbits, and follow the bifurcation structure of their solution branches under parametric continuation in Rayleigh number. We discover bifurcation scenarios involving the simultaneous creation of five non-symmetrically-related branches emerging from transverse rolls. The existence of these branches follows from the breaking of the spatial $D_4$ and $O(2)$ symmetries in a Hopf bifurcation. A heteroclinic cycle between two unstable periodic orbits emerges immediately after the Hopf bifurcation point. Additional unstable periodic orbits capturing both large- and small-scale features of the SDP pattern are also identified, and their global bifurcation is revealed.

\par The rest of the manuscript is structured as follows. In \S \ref{SDP_system_method}, we present the numerical methods used in our research. The analysis of direct numerical simulations in both large and small spatial domains are shown in \S \ref{SDP_large_simulation}. In \S \ref{SDP_TO1TO2}, we discuss the linear instability of the secondary state as well as its simultaneous bifurcations to the tertiary states. In \S \ref{SDP_orbit_SDP}, a global bifurcation of a relevant periodic orbit is presented. In \S \ref{SDP_edgestate}, we investigate the edge state between two periodic orbits exhibiting the same spatial symmetry. Eventually, we conclude with a summary of key findings and a discussion in \S \ref{SDP_conclusion}.

\section{System and numerical concepts}
\label{SDP_system_method}
\par The numerical methods used in this work are the same as those described in \citet{Reetz2020a, Reetz2020, Zheng2024part1, Zheng2024part2, Ringenbach2025} and \citet{Zheng2025PhD}. Therefore, we refer the reader to these publications for more details and provide a succinct overview in this section.

\begin{figure}
    \centering
    \includegraphics[width=0.7\columnwidth]{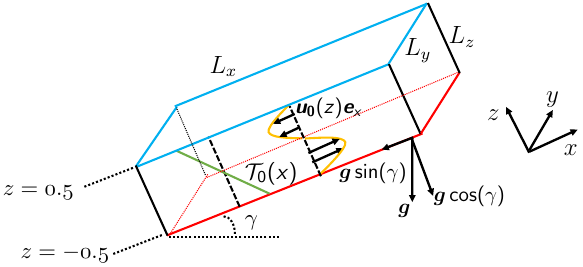}
    \captionsetup{font={footnotesize}}
    \captionsetup{width=13.5cm}
    \captionsetup{format=plain, justification=justified}
    \caption{\label{SDP_ILC_figure} Schematic of the convection cell where the confined fluid layer is inclined against the gravity by angle $\gamma$. The co-ordinates $x$, $y$ and $z$ represent the streamwise, spanwise and wall-normal directions, respectively. The flow is bounded between two fixed walls in $z$, at $z=-0.5$ where the fluid is heated and at $z=0.5$ where the fluid is cooled. For $\gamma > 90 \degree$, the fluid is heated from above. The velocity ($\boldsymbol u_0$) and temperature ($\mathcal{T}_0$) profile of laminar base solution are shown by orange curve and green line, respectively. All flow field snapshots presented in this paper are temperature field visualised at the midplane $z=0$.}
\end{figure}

\par We perform direct numerical simulations (DNS) by using the ILC extension module of the pseudo-spectral code \textit{Channelflow 2.0} \citep{Gibson2019}. The Channelflow-ILC code time-integrates the non-dimensionalised Oberbeck--Boussinesq equations
\begin{align}
    \dfrac{\partial \boldsymbol u}{\partial t}+ (\boldsymbol u \cdot \nabla)\boldsymbol u &= -\nabla p + \sqrt{\frac{Pr}{Ra}} \nabla^2 \boldsymbol u - \hat{\boldsymbol{g}}\mathcal{T}, \label{SDP_appa} \\
    \dfrac{\partial \mathcal{T}}{\partial t} + (\boldsymbol u \cdot \nabla)\mathcal{T} &= \frac{1}{\sqrt{PrRa}}\nabla^2 \mathcal{T}, \label{SDP_appb} \\
    \nabla \cdot  \boldsymbol u&= 0, \label{SDP_appc} 
\end{align}
in a channel geometry depicted in figure \ref{SDP_ILC_figure}. In \eqref{SDP_appa}--\eqref{SDP_appc}, $\boldsymbol u$, $\mathcal{T}$ and $p$ stand for total velocity, total temperature, and pressure relative to the hydrostatic pressure, respectively. The flow is subject to periodic boundary conditions in the streamwise $x$ and spanwise $y$ directions, and a zero mean pressure gradient integral constraint is imposed in $x$ and $y$. The computational domain is bounded at $z=\pm 0.5$ where $\boldsymbol u(z=\pm0.5)=\mathbf{0}$ and $\mathcal{T}(z=\pm0.5)=\mp0.5$. The domain is inclined against gravity by an angle $\gamma$, resulting in the buoyancy force contributing to both the $x$- and $z$-momentum equations:
\begin{equation}
    \hat{\boldsymbol{g}}= -\sin(\gamma)\boldsymbol e_x - \cos(\gamma)\boldsymbol e_z.
\end{equation}
Equations \eqref{SDP_appa}--\eqref{SDP_appc}, together with the given boundary conditions, admit the following laminar solution: 
\begin{align}
    & \boldsymbol u_0(z) = \dfrac {\sin(\gamma)\sqrt{Ra}}{6\sqrt{Pr}} \left( z^3 - \frac{1}{4}z \right) \boldsymbol e_x,	\label{SDP_laminara} \\
    &\mathcal{T}_0(z) = -z,  \label{SDP_laminarb} \\
    &p_0(z) = \Pi - \cos(\gamma)z^2/2,	\label{SDP_laminarc} 
\end{align}
with $\Pi$ being an arbitrary pressure constant. The laminar velocity and temperature profiles are sketched in figure \ref{SDP_ILC_figure}.

\par The ILC system is equivariant under reflection with respect to the $y$ axis (\ref{SDP_sym_b}) or the $x$--$z$ plane (\ref{SDP_sym_c}) and under translation in the $x$ and $y$ directions (\ref{SDP_sym_a}):
\begin{subequations}
    \begin{eqnarray}
        &\pi_y[U,V,W,\mathcal{T}](x,y,z) \equiv [U,-V,W,\mathcal{T}](x, -y,z), \label{SDP_sym_b}\\
        &\pi_{xz}[U,V,W,\mathcal{T}](x,y,z) \equiv [-U,V,-W,-\mathcal{T}](-x, y,-z), \label{SDP_sym_c} \\
        &\tau(\Delta x, \Delta y)[U,V,W,\mathcal{T}](x,y,z) \equiv [U,V,W,\mathcal{T}](x+\Delta x, y + \Delta y,z). \label{SDP_sym_a}
    \end{eqnarray}
\end{subequations}
These symmetry operations form the equivariance group of the ILC system, denoted by $S_{ILC} \equiv \braket{\pi_{xz}, \pi_y, \tau(\Delta x, \Delta y)} \sim [O(2)]_{xz} \times [O(2)]_y$, where $\braket{}$ implies all products of the elements given in the brackets.

\par We use two domain sizes in this work: $[L_x, L_y, L_z] = [26.65, 12.1, 1]$ and $[100, 50, 1]$, that are discretised using $[N_x, N_y, N_z] = [96, 48, 31]$ and $[384, 192, 31]$ Fourier–Fourier-Chebychev collocation points, respectively. The domain $[L_x, L_y, L_z] = [26.65, 12.1, 1]$ has been used by \citet{Reetz2020a} and \citet{Reetz2020} for bifurcation analysis; this domain size has been chosen based on the SDP pattern wavelength suggested by linear stability analysis \citep{Subramanian2016}. The Rayleigh number is the sole control parameter in this work, with $\gamma$ and $Pr$ fixed to be $100\degree$ and $1.07$, respectively. We will occasionally use the rescaled Rayleigh number $\epsilon = (Ra - Ra_{c})/Ra_{c}$, where $Ra_{c}=9122$ is the critical Rayleigh number corresponding to the onset of convection at $(\gamma, Pr)=(100\degree,1.07)$ in the domain $[L_x, L_y, L_z] = [26.65, 12.1, 1]$.

\par Invariant solutions, including equilibria and periodic orbits, are numerically computed using the standard Newton-based shooting method, which identifies the state vector $\boldsymbol{x}^{*}(t)$ satisfying the recurrence equation
\begin{equation}
\label{SDP_eq:recurrence}
    \mathcal{G}(\boldsymbol{x}^{*})=\sigma \mathcal{F}^T(\boldsymbol{x}^{*}) - \boldsymbol{x}^{*} = 0,
\end{equation}
where $\mathcal{F}^T$ is the evolution operator corresponding to the time integration of \eqref{SDP_appa}--\eqref{SDP_appc} from an initial state $\boldsymbol{x}^{*}$ over a finite time period $T$, and $\sigma$ is a symmetry operator. The time interval $T$ is arbitrary for a steady solution, and is the period of a time-periodic solution. To construct invariant solutions underlying the SDP pattern, we extract initial guesses from DNS. These guesses are subsequently converged to an invariant solution by solving the root-finding problem \eqref{SDP_eq:recurrence} using Newton iterations combined with the hookstep trust-region optimisation \citep{Viswanath2009}. For alternative methods of computing unstable invariant solutions, see, for example \cite{Farazmand2016, Azimi2022, Parker2022, Ashtari2023}. For each converged state, linear stability is analysed by computing the leading eigenvalues of the linearised dynamics using the Arnoldi algorithm, and bifurcations of the invariant solutions are tracked using parametric continuations in Rayleigh number.

\par We employ the $L_2$-norm of the temperature deviations $\theta \equiv \mathcal{T} - \mathcal{T}_0$ as an observable for plotting bifurcation diagrams:
\begin{equation}
    \|\theta \|_2 = \left( \frac{1}{L_x}\frac{1}{L_y}\int_{0}^{L_x} \int_{0}^{L_y} \int_{-0.5}^{0.5} \theta^2(x, y, z)\, dzdydx \right)^{\frac{1}{2}}. 
\end{equation}
The branches of equilibria are represented by curves showing $\| \theta \|_2$ as a function of $Ra$. For periodic orbits, the maximum and minimum of $\| \theta \|_2$ along an orbit are plotted. In addition, we visualise the phase portrait using 3D projections onto the rate of thermal energy input $I$ due to buoyancy forces, the viscous dissipation rate $D$, and the space-averaged streamwise velocity deviation $u_{\text{bulk}}$. Throughout the paper, flow fields and eigenvectors are visualised using the temperature deviation field $\theta$ at the midplane $z=0$.

\section{Direct numerical simulations of switching diamond panes patterns}
\label{SDP_large_simulation}
\begin{figure}
    \centering
    \includegraphics[width=\columnwidth]{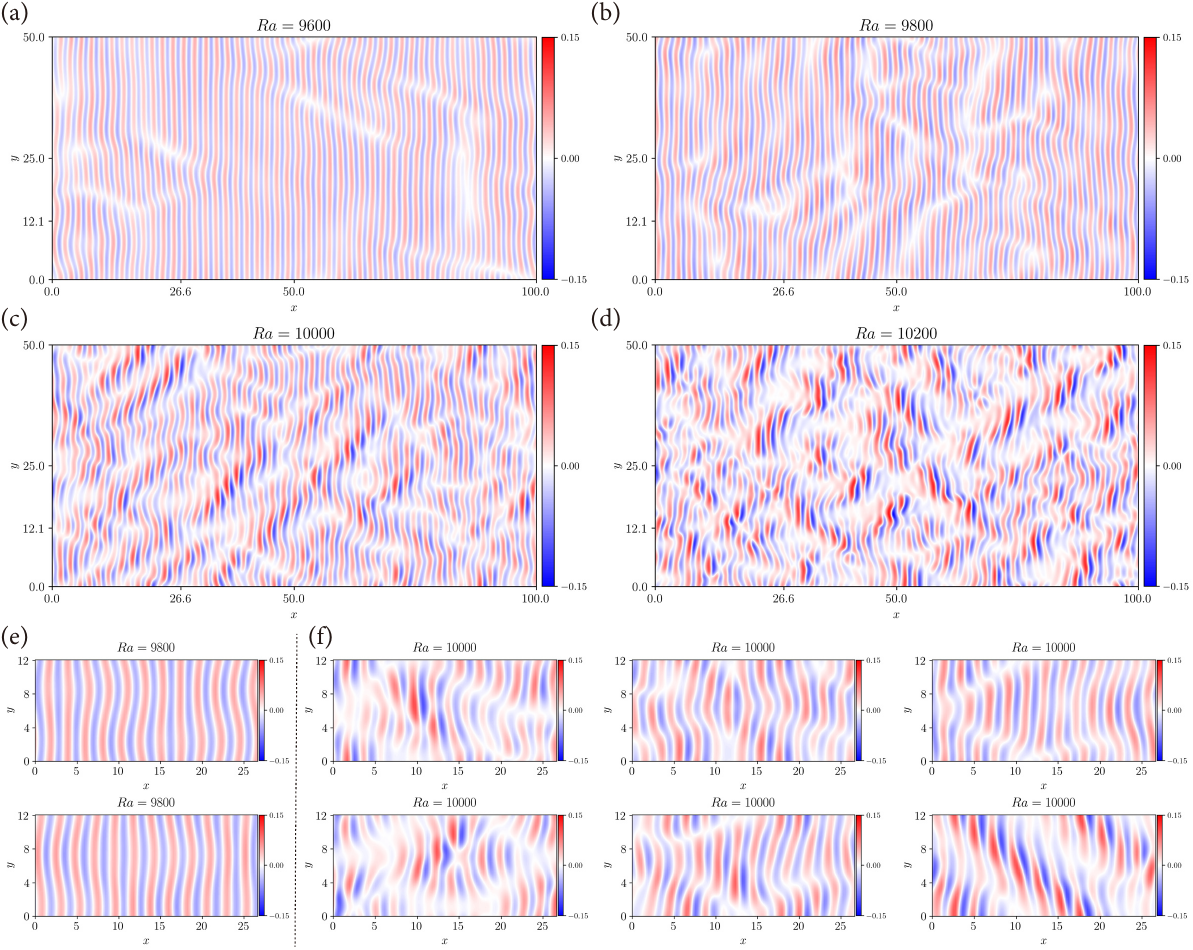}
    \captionsetup{font={footnotesize}}
    \captionsetup{width=13.5cm}
    \captionsetup{format=plain, justification=justified}
    \caption{\label{SDP_large_domain} Snapshots of the midplane temperature field. (a--d) Simulations in the large spatial domain $[L_x, L_y, L_z] = [100, 50, 1]$ at four Rayleigh numbers $Ra=9600$ (a), $9800$ (b), $10000$ (c), and $10200$ (d). From (a) to (d), the flow transitions from almost straight (in $y$) transverse convection rolls to spatio-temporally chaotic switching of large amplitude regions of transverse rolls. The complex pattern in (b) and (c) is called switching diamond panes (SDP). (e--f) Simulations in the small periodic domain $[L_x, L_y, L_z] = [26.6, 12.1, 1]$ at $Ra=9800$ (e, two snapshots) and $Ra=10000$ (f, six snapshots). The time series of $\|\theta\|_2$ at $Ra=9800$ and $10000$ are shown in panels (b) and (f) of figure \ref{SDP_timeseries9750_10000}, respectively. The same colour bar is used in all plots.}
\end{figure}

\par We perform DNS in two domains $[L_x, L_y, L_z] = [26.65, 12.1, 1]$ and $[100, 50, 1]$, to gain insight into the spatio-temporal dynamics of the SDP pattern and its characteristic features. These two domains are hereafter referred to as the small and large domains, respectively.

\subsection{Dynamics in the large (spatially extended) domain}
\par Representative snapshots from DNS at four successively increasing Rayleigh numbers are shown in figures \ref{SDP_large_domain}(a--d). In figure \ref{SDP_large_domain}(a) at $Ra=9600$, we observe approximately 45 transverse convection rolls which are almost straight in the spanwise ($y$) direction. In addition to these transverse rolls, few low-amplitude defects inducing large-wavelength stripes are observed. Most of these stripes are obliquely oriented and are therefore referred to as oblique stripes; however, alignment with the $y$-axis is also possible. In figure \ref{SDP_large_domain}(b) at $Ra=9800$, the modulation of transverse rolls is more pronounced, and a larger number of oblique stripes and defect regions are present. According to \citet{Daniels2000}, $Ra=9800$ ($\epsilon \approx 0.07$) corresponds approximately to the onset of the SDP pattern.

\par At $Ra=10000$ ($\epsilon \approx 0.1$) in figure \ref{SDP_large_domain}(c), transverse rolls exhibit more chaotic dynamics; the defects and oblique stripes continue to spread throughout the entire domain. Furthermore, some high-amplitude regions with oblique orientation are observed, which are almost parallel to each other. We refer to these structures as oblique bursts.  The interlaced and intersected oblique stripes and bursts form the shape of diamond panes which appear, disappear, and then reappear in an erratic manner. This complex pattern is called switching diamond panes; see and compare with the shadowgraph images in figure 9(a) of \citet{Daniels2000}. Further increasing the thermal driving to $Ra=10200$, the oblique bursting spots spread over the domain with an increased switching frequency and roll intensity. This triggers the transition from the SDP pattern to the ``longitudinal bursts within diamond panes'' pattern \citep{Daniels2000}, the investigation of which lies beyond the scope of this paper.

\subsection{Dynamics in the small domain}
\par Snapshots from simulations in the small domain at $Ra=9800$ and $10000$ are shown in figures \ref{SDP_large_domain}(e,f). We have focused on these two Rayleigh numbers because the SDP pattern emerges within this $Ra$-range. At $Ra=9800$, the dynamics consists of low-amplitude oscillations of transverse rolls, without any defects and large-scale stripes. This observation might be surprising at first glance, when comparing to large-domain simulations at the same $Ra$. We will discuss in detail in \S \ref{SDP_hetero_PO1PO3} that the dynamics at $Ra=9800$ in the small domain consists of a heteroclinic cycle between two unstable periodic orbits. Due to the existence of this attracting heteroclinic cycle, the complex dynamics is effectively simplified to visiting these two unstable orbits. The heteroclinic cycle is unstable in the large domain and thus overwhelmed by the defects and large-scale structures. We have also confirmed that simulations in the small domain starting from random initial conditions (amplitude noise) indeed pass through transients capturing stripes and defects (not shown) before settling onto the dynamically stable heteroclinic cycle.

\par At $Ra=10000$, complex spatio-temporal dynamics is observed, and we have shown six characteristic snapshots in figure \ref{SDP_large_domain}(f). Each snapshot corresponds to an instantaneous temperature field in the simulation, and all six can be approximately identified at different locations in the large-domain snapshot (figure \ref{SDP_large_domain}c) at a given time. At this $Ra$, the heteroclinic cycle mentioned above has broken, and so chaotic dynamics takes place. Snapshots in figure \ref{SDP_large_domain}(f) capture defects, stripes, bursts, and diamond-shaped modulations of rolls. The states in these snapshots are not exactly symmetric, but most of them exhibit visual features reminiscent of reflection and/or translation symmetries. We will converge equilibria, travelling waves, and periodic orbits that exhibit characteristics very similar to these snapshots and are exactly symmetric, as imposed numerically. For the rest of the paper, we study the dynamics and invariant states in the small domain, $[L_x, L_y, L_z] = [26.65, 12.1, 1]$.

\section{Local supercritical Hopf bifurcations from transverse rolls}
\label{SDP_TO1TO2}
\begin{figure}
    \centering
    \includegraphics[width=\columnwidth]{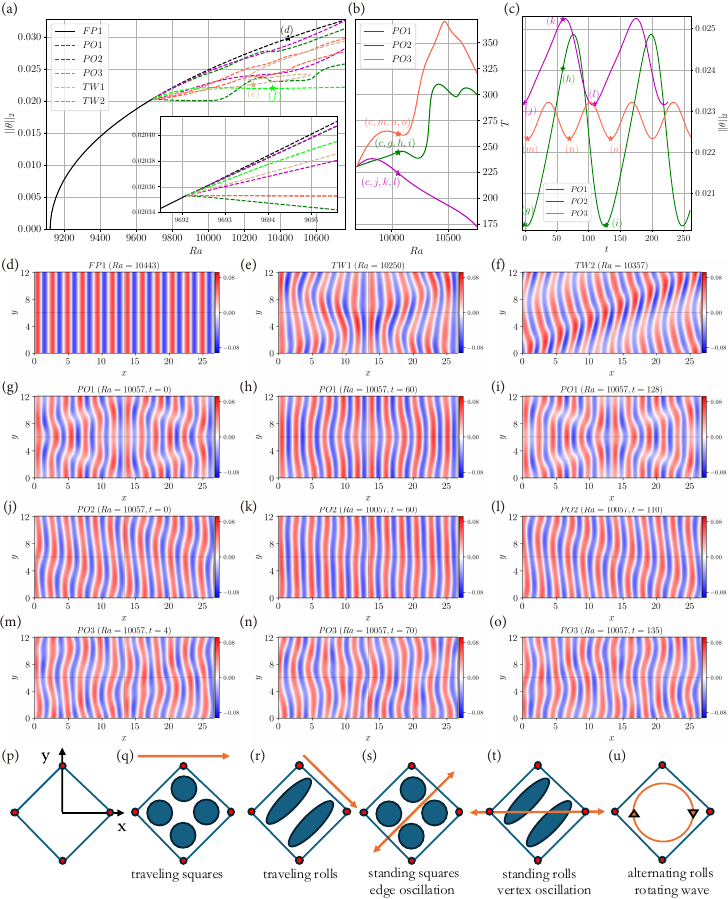}
    \captionsetup{font={footnotesize}}
    \captionsetup{width=13.5cm}
    \captionsetup{format=plain, justification=justified}
    \caption{\label{SDP_TRTO_BD} (a) Bifurcation diagram of one equilibrium state, two travelling waves, and three periodic orbits. Solid and dashed curves indicate linearly stable and unstable states, respectively. The inset zooms in on the Hopf bifurcation at which five solution branches---TW1, TW2, PO1, PO2, and PO3---bifurcate simultaneously from FP1. The two curves representing the maximum and minimum of $\| \theta \|_2$ for PO3 (shown in orangered) are too close to be distinguished near the bifurcation point. (b) Periods of the three periodic orbits. (c) Time series of $\|\theta\|_2$ for the three periodic orbits at $Ra=10057$. (d--o) Snapshots of the midplane temperature of FP1, TW1, TW2, PO1, PO2, and PO3. In (a--c), the stars and labels indicate the locations of the corresponding states visualised in (d--o). The same colour  bar is used in all plots. (q--u) Schematics of five states simultaneously bifurcating from a square lattice (p) in a Hopf bifurcation. The patterns in (q,r) travel in the direction of the orange arrows; in (s,t), they oscillate between opposite edges and vertices of the square; and in (u), they rotate around the origin. The names describing each of the states in (q--u) are those used by \citet{Silber1991} and \citet{Swift1988}. Schematics are inspired by figure 2 of \citet{Swift1988} and figure 4 of \citet{Silber1991}.}
\end{figure}

\par To understand the complex SDP patterns seen in figure \ref{SDP_large_domain}, a natural step is to investigate the linear stability of the secondary state, transverse rolls. Figure \ref{SDP_TRTO_BD}(a) shows the bifurcation diagram of the identified invariant solutions over a range of $Ra$ relevant to the observations discussed in \S \ref{SDP_large_simulation}. Transverse rolls, denoted with FP1 in the following, bifurcate from the laminar base flow at $Ra=9122$. The equilibrium state FP1, as shown in figure \ref{SDP_TRTO_BD}(d), consists of 12 identical straight rolls and has reflection and translation symmetries
\begin{equation}
    \label{SDP_eq:symmetries_of_FP1}
    S_{\text{FP1}} \equiv \braket{\pi_{xz},\pi_y,\tau(L_x/12,\Delta y)} \sim [D_{12}]_{xz} \times [O(2)]_y.
\end{equation}
Equilibrium FP1 loses linear stability at $Ra=9692.2$, which is emphasised in the inset of figure \ref{SDP_TRTO_BD}(a). At $Ra=9692.2$, eight eigenvalues---four complex conjugate pairs with the same oscillation frequency $\omega = 0.0273$---cross simultaneously the imaginary axis; the centre manifold at the bifurcation point is thus eight-dimensional. We identified three standing waves and two travelling waves, all bifurcating from FP1 at $Ra=9692.2$ in supercritical Hopf bifurcations. This section is devoted to explaining these non-generic simultaneous bifurcations and the related eight-dimensional neutral eigenspace.

\subsection{Three standing waves and two travelling waves}
\label{SDP_PO1PO2PO3TW1TW2}
\par We will start by discussing each of the solution branches that we identified, and particularly the symmetries of each state. 

\subsubsection{States PO1 and TW1}
\label{SDP_PO1TW1}
\par \citet{Reetz2020} reported a periodic orbit bifurcating from FP1 and called it transverse oscillations, a name also used by \citet{Subramanian2016}. This orbit has been recomputed here and is referred to as PO1. The bifurcation diagram of PO1 is shown in figure \ref{SDP_TRTO_BD}(a), its periods in figure \ref{SDP_TRTO_BD}(b), the time series of its temperature norm at $Ra=10057$ in figure \ref{SDP_TRTO_BD}(c), and selected snapshots in figures \ref{SDP_TRTO_BD}(g--i). Orbit PO1 has the spatial symmetries $S_{\text{PO1}} \equiv \braket{\pi_{xz}, \pi_y, \tau(L_x/2, L_y/2)}$, as well as the spatio-temporal symmetry:
\begin{align}
    (u,v,w,\theta)(x,y,z,t+T/2) = (u,v,w,\theta)(x+L_x/2,y,z,t),
    \label{SDP_spaiotempo_PO1}
\end{align}
where $T$ is the period of PO1 at a given $Ra$. This spatio-temporal symmetry can be seen by comparing, for instance, figures \ref{SDP_TRTO_BD}(g) and (i). Orbit PO1 is a pre-periodic orbit and a standing wave solution, as suggested by \citet{Reetz2020}. Within the subspace of flow fields symmetric under $S_{\text{PO1}}$, PO1 is linearly stable for $9692.2<Ra \lesssim 10100$. However, when perturbations that break the spatial symmetries of PO1 are allowed, this periodic orbit becomes unstable at onset. We have continued PO1 until $Ra=11000$.

\par The travelling wave paired with PO1 is TW1, shown in figure \ref{SDP_TRTO_BD}(e). State TW1 has the spatial symmetries $S_{\text{TW1}} \equiv \braket{\pi_{xz}, \tau(L_x/2, L_y/2)}$ and travels in the $y$ direction with speed $c=L_y/T=12.1/230=0.0526$ where $T$ is the period of PO1 at the bifurcation point. State TW1 is unstable (from its creation) within the symmetry subspace $S_{\text{TW1}}$ and thus more unstable in the full phase space. State TW1 undergoes several saddle--node bifurcations between $10200\lesssim Ra\lesssim 10600$, and we have continued it until $Ra=10800$. 

\subsubsection{States PO2 and TW2}
\par We have found another periodic orbit, denoted by PO2 and shown in figures \ref{SDP_TRTO_BD}(j--l); see also figures \ref{SDP_TRTO_BD}(a--c). Orbit PO2 has the spatial symmetries $S_{\text{PO2}} \equiv \braket{\pi_{xz}\pi_y, \tau(L_x/12, -L_y/12)}$, as well as the spatio-temporal symmetry \eqref{SDP_spaiotempo_PO1}. Similar to PO1, PO2 is also a pre-periodic orbit and a standing wave. Orbit PO2 is unstable within $S_{\text{PO2}}$.

\par The second travelling wave we found is TW2 (paired with PO2), shown in figure \ref{SDP_TRTO_BD}(f). State TW2 has the spatial symmetries $S_{\text{TW2}} \equiv \braket{\tau(L_x/12, L_y/12)}$, and travels in the $y$ direction. State TW2 is unstable within $S_{\text{TW2}}$ and is continued until $Ra=11000$. The existence of travelling wave solutions may explain the observed travelling dynamics of the SDP pattern.

\subsubsection{State PO3}
\par The third periodic orbit that we found is PO3, shown in figures \ref{SDP_TRTO_BD}(m--o); see also figures \ref{SDP_TRTO_BD}(a--c). The oscillation amplitude of PO3 is smaller than that of PO1 and PO2, thus the maxima and minima of $\| \theta \|_2$ of PO3 almost coincide close to the bifurcation point (see the inset of figure \ref{SDP_TRTO_BD}a). Orbit PO3 has the spatial symmetries $S_{\text{PO3}} \equiv \braket{\tau(L_x/2, L_y/2)}$, as well as the spatio-temporal symmetry
\begin{align}
    (u,v,w,\theta)(x,y,z,t+T/4) &= \pi_y(u,v,w,\theta)(x-L_x/4,y,z,t), \label{SDP_spaiotempo1_PO3}
\end{align}
where $T$ is the period of PO3. This spatio-temporal symmetry can be seen by comparing, for instance, figures \ref{SDP_TRTO_BD}(m,n). Note that applying \eqref{SDP_spaiotempo1_PO3} twice, the initial and final states are related by $\sigma \equiv \tau(\pm L_x/2,0)$; compare figures \ref{SDP_TRTO_BD}(m,o). Orbit PO3 is also a pre-periodic orbit and a standing wave, but it does not have a paired travelling wave; this will be explained in \S \ref{SDP_simultaneous_bif}. Orbit PO3 is unstable within $S_{\text{PO3}}$, and we have managed to continue it until $Ra=10800$. Note that close to the bifurcation of PO1, PO2, and PO3 from FP1, the period $T\approx230$ of PO1--PO3 closely matches $2\pi/\omega$, as is typical for Hopf bifurcations.

\subsection{Simultaneous Hopf bifurcations}
\label{SDP_simultaneous_bif}
\par As is seen in \S \ref{SDP_PO1PO2PO3TW1TW2}, even though five solution branches bifurcate together from FP1, they differ qualitatively and quantitatively; they have different symmetries, temperature norms $\| \theta \|_2$, and periods (for orbits). This subsection will address the theoretical basis for these simultaneous bifurcations.

\subsubsection{Equivariant bifurcation theory}
\par It is known that Hopf bifurcations breaking the $O(2)$ symmetry of an equilibrium state give rise simultaneously to a standing wave branch and a travelling wave branch. The travelling wave branch includes two physically identical solutions that travel with the same speed but in opposite directions---the left- and right-travelling waves. Close to the Hopf bifurcation point, the standing wave can be interpreted as the superposition of these two counter-propagating travelling waves. Such bifurcations involve a four-dimensional neutral eigenspace. The corresponding normal form---namely ordinary differential equations for the amplitudes of the critical modes---is analysed in \citet{Golubitsky1985} and \citet{Knobloch1986}, and examples in high-dimensional hydrodynamic systems can be found in \citet{Boronska2006} and \citet{Reetz2020}. In our case, the existence of both TW1-PO1 and TW2-PO2 pairs can be understood as a consequence of breaking the $[O(2)]_y$ symmetry of FP1 (see \eqref{SDP_eq:symmetries_of_FP1}). However, two key observations remain to be understood. First, not one but two pairs bifurcate simultaneously, and second, PO3 bifurcates together with them from the same parent branch. To the best of our knowledge, this bifurcation scenario has not been previously reported in 3D Navier--Stokes systems. To understand these simultaneous bifurcations, we briefly review the theory of pitchfork and Hopf bifurcations that break the $D_4$ symmetry of a steady state.

\par In pitchfork bifurcations that break the $D_4$ symmetry of an equilibrium, two purely real eigenvalues cross the imaginary axis at the same time. This steady-state bifurcation gives rise to two different solution branches simultaneously, as described by the corresponding normal form in \citet{swift1985bifurcation} and \citet{Hoyle2006}. In many hydrodynamic configurations, one of the two bifurcating branches corresponds to non-linear states with diagonally-oriented coherent features, referred to as diagonal states or rolls. The other branch consists of non-linear states exhibiting features which are aligned with a periodic direction, known as rectangular states or squares. Examples of such simultaneous bifurcations in 3D hydrodynamic systems can be found in \citet{bergeon2001three, bengana2019spirals, Reetz2020, Zheng2024part1} and \citet{Zheng2024part2}.

\begin{table}
    \centering
    \begin{tabular}{ccccc}
        Solution & \cite{Swift1988} & \cite{Silber1991} & \cite{Rucklidge1997} \\[0.2cm]
        PO1 & edge oscillation & standing squares (SS) & diagonally pulsating squares \\
        TW1 & -- & travelling squares (TS) & -- \\
        PO2 & vertex oscillation & standing rolls (SR) & pulsating squares \\
        TW2 & -- & travelling rolls (TR) & -- \\
        PO3 & rotating wave & alternating rolls (AR) & alternating pulsating waves \\
        S6 & non-symmetric periodic solution & standing cross-rolls (SCR) & -- \\
    \end{tabular}
    \captionsetup{font={footnotesize}}
    \captionsetup{width=13.5cm}
    \captionsetup{format=plain, justification=justified}
    \caption{Correspondence of five simultaneously bifurcating states (that are present in our configuration) with the equivariant bifurcation theory discussed in \citet{Swift1988, Silber1991} and \citet{Rucklidge1997}. A sixth solution (periodic orbit, S6) is predicted by the theory but cannot exist in our case, where the five states all bifurcate supercritically with none being linearly stable (see text). States not reported are indicated by ``--''. The abbreviations used by \citet{Silber1991} are given in parenthesis, and the schematics for SS, TS, SR, TR, and AR are shown in figures \ref{SDP_TRTO_BD}(q--u).}
    \label{SDP_Hopf_comparison_literature}
\end{table}

\par Hopf bifurcations breaking the $D_4$ symmetry of an equilibrium state are studied by \citet{Golubitsky1986, Swift1988, Silber1991} and \citet{Rucklidge1997}. \citet{Swift1988} derived and analysed the so-called truncated Birkhoff normal form. In this analysis, the continuous translational symmetry along two orthonormal directions of the square is not included, thus the resulting states cannot drift or travel---no travelling waves allowed. \citet{Swift1988} reported that at least three different oscillatory solutions bifurcate simultaneously from the trivial solution; these are called edge oscillation, vertex oscillation, rotating wave, and possibly a fourth non-symmetric periodic solution; see schematics in figure 2 of \citet{Swift1988} and figures \ref{SDP_TRTO_BD}(s--u). In the edge and vertex oscillations, the ``point particle'' oscillates along a line joining opposite edges or vertices of a square (figures \ref{SDP_TRTO_BD}s and \ref{SDP_TRTO_BD}t); in the rotating wave, the point particle follows a circular closed trajectory with square symmetry (figure \ref{SDP_TRTO_BD}u). These periodic solutions were later discussed in \citet{Silber1991} and \citet{Rucklidge1997} who used slightly different names; see a summary of these names in table \ref{SDP_Hopf_comparison_literature}. 

\par \citet{Silber1991} extended the work of \citet{Swift1988} by imposing periodic boundary conditions in the two orthogonal directions and thereby incorporating the spatial translation symmetry of the square; this allows travelling waves to exist. \citet{Silber1991} found that generically there are at least five solution branches---two travelling waves and three standing waves---bifurcating simultaneously from the trivial solution, in the case of the cubic truncation of the normal form. These five solutions are guaranteed by the equivariant Hopf theorem \citep{Golubitsky1985}. There is potentially a sixth solution which is possibly the fourth (non-symmetric periodic) solution in \citet{Swift1988}; this solution is a standing wave, is always unstable, and its existence depends on the cubic coefficients of normal form. The two new travelling waves (in addition to \citet{Swift1988}) are called travelling squares and travelling rolls; see the shadowgraph images of a convecting fluid layer in figure 4 of \citet{Silber1991} and schematics in figures \ref{SDP_TRTO_BD}(q--r). In figures \ref{SDP_TRTO_BD}(q--t), we use two ovals in a square to represent travelling and standing rolls, and four circles in a square to represent travelling and standing squares; these follow the conventional pictures of rolls and squares in steady-state bifurcations from $D_4$-symmetric states. \citet{Silber1991} computed the stability of these five solutions and reported 34 possible bifurcation diagrams. In particular, it is possible for all five branches to bifurcate supercritically with none being stable; this corresponds to cases 12 and 34 of figure 9 in \citet{Silber1991}. In case 12, the standing squares (SS) has to be unstable, and in case 34, the standing rolls (SR) has be be unstable. In both cases, one is necessarily in a part of the (normal form) coefficient space where the sixth solution standing cross-rolls (SCR) does not exist; see figure 5 of \citet{Silber1991}.

\subsubsection{Our hydrodynamic system}
\par The five simultaneously bifurcating branches (PO1, PO2, PO3, TW1, and TW2) discussed in \S \ref{SDP_PO1PO2PO3TW1TW2} reflect precisely the equivariant bifurcation theory described in \citet{Silber1991}. In our hydrodynamic configuration, the continuous translation symmetry (of the square) is analogous to the $O(2)$ symmetry of FP1 in the $y$-direction, allowing travelling waves in $y$ to exist. The $[D_4]_{xz}$ symmetry is embedded in the $[D_{12}]_{xz}$ symmetry of FP1; note that, if a state has $D_{12}$ symmetry, it necessarily has $D_{2}$, $D_{3}$, $D_{4}$, and $D_{6}$ symmetries. Each of the five bifurcating states breaks the $D_4$ symmetry of FP1 in the $xz$-direction at the same time as it breaks the $O(2)$ symmetry in the $y$-direction. 

\par Combining figures \ref{SDP_TRTO_BD}(d--l) and spatial symmetries of these invariant solutions, it can be inferred that PO1 (with $y$- and $xz$-reflection) and TW1 (with $xz$-reflection) are rectangular states; they correspond to standing and travelling squares in \citet{Silber1991}. Orbit PO2 (with $\tau(L_x/12, -L_y/12)$ translation) and TW2 (with $\tau(L_x/12, L_y/12)$ translation) are diagonal states; they correspond to standing and travelling rolls in \citet{Silber1991}. For standing squares (edge oscillation, figure \ref{SDP_TRTO_BD}s) and standing rolls (vertex oscillation, figure \ref{SDP_TRTO_BD}t), one complete oscillation consists of two half-oscillations related by spatio-temporal symmetries. This is exactly the case for PO1, PO2, and their spatio-temporal symmetry \eqref{SDP_spaiotempo_PO1} after a half-period. Orbit PO3 (with $\tau(L_x/2, L_y/2)$ translation) corresponds to the alternating rolls in \citet{Silber1991}. In a square, alternating rolls change orientation by $90\degree$ after a quarter of the period, and return to the original state after a full period; see figure 4 of \citet{Silber1991} and figure \ref{SDP_TRTO_BD}(u). This is precisely the behaviour seen for PO3; it has the spatio-temporal symmetry \eqref{SDP_spaiotempo1_PO3} after a quarter-period. Orbit PO3 does not have a paired travelling wave, as is the case in the normal-form analysis. The correspondence between the five solution branches and the theory is summarised in table \ref{SDP_Hopf_comparison_literature}. In addition, all five branches bifurcate supercritically from FP1, and they are all linearly unstable directly after their creation, as discussed in \S \ref{SDP_PO1PO2PO3TW1TW2}. This suggests that the potential sixth solution (SCR) predicted by \citet{Silber1991} does not exist in this case.

\begin{figure}
    \centering
    \includegraphics[width=\columnwidth]{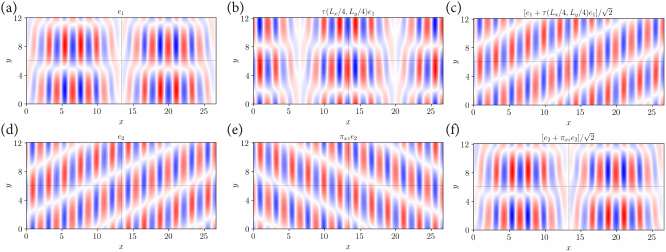}
    \captionsetup{font={footnotesize}}
    \captionsetup{width=13.5cm}
    \captionsetup{format=plain, justification=justified}
    \caption{\label{SDP_eigen_D4} (a) Eigenvector $e_1$ responsible for FP1$\rightarrow$TW1 bifurcation and (b) its quarter-diagonal translation $\tau(L_x/4,L_y/4) e_1$. (c) Linear combination $(e_1 + \tau(L_x/4,L_y/4) e_1)/\sqrt{2}$. (d) Eigenvector $e_2$ responsible for FP1$\rightarrow$TW2 bifurcation and (e) its $xz$-reflection $\pi_{xz} e_2$. (f) Linear combination $(e_2 + \pi_{xz} e_2)/\sqrt{2}$. All eigenvectors are visualised using the temperature field in the $x$--$y$ plane at $z=0$. The same colour bar is used in all plots.}
\end{figure}

\par To make a connection with the breaking of $D_4$-symmetry in pitchfork bifurcations, we show in figures \ref{SDP_eigen_D4}(a) and (d), the eigenmodes of FP1 that give rise to TW1 and TW2 (with a fixed phase), respectively; we denote these by $e_{1}$ and $e_2$. The eigenmodes $e_{1,2}$ are obtained by subtracting FP1 from TW1 and TW2 at $Ra=9693$; this enables us to choose the correct vector from the span of the neutral eigenvectors of FP1, returned by the Arnoldi method at the bifurcation point. We have verified the dimension of the centre manifold calculated numerically by confirming that $e_{1}$ and $e_2$ can be expressed as linear combinations of the eigenmodes returned by the Arnoldi method. Figures \ref{SDP_eigen_D4}(b) and (e) show the shifted and reflected versions of $e_{1,2}$, and figures \ref{SDP_eigen_D4}(c) and (f) show the equal superposition of the two symmetry-related eigenmodes. We find that figure \ref{SDP_eigen_D4}(c) is identical to figure \ref{SDP_eigen_D4}(d), and figure \ref{SDP_eigen_D4}(f) is the same as figure \ref{SDP_eigen_D4}(a). This is a manifestation of the fact that linear combinations of two symmetry-related diagonal (rectangular) states give the rectangular (diagonal) states, as in the normal form of pitchfork bifurcations with $D_4$ symmetry. The same analysis could be carried out for the eigenmodes leading to PO1 and PO2, and gives the same results (not shown).

\par Note that FP1 also has $D_{6}$ (hexagonal) symmetry. Hopf bifurcation from $D_6$-symmetric states and its normal form are discussed in \citet{Roberts1986}, who report that the $D_{6}$ symmetry forces the dimension of the eigenspace to be a multiple of 12, and there are at least 11 different types of bifurcating solutions, including a variety of standing and travelling waves. However, in the present case, the $D_{6}$ symmetry of FP1 does not play a role in the observed bifurcation scenarios.

\subsection{Heteroclinic cycle between two unstable periodic orbits (PO1 and PO3)}
\label{SDP_hetero_PO1PO3}
\begin{figure}
    \centering
    \includegraphics[width=\columnwidth]{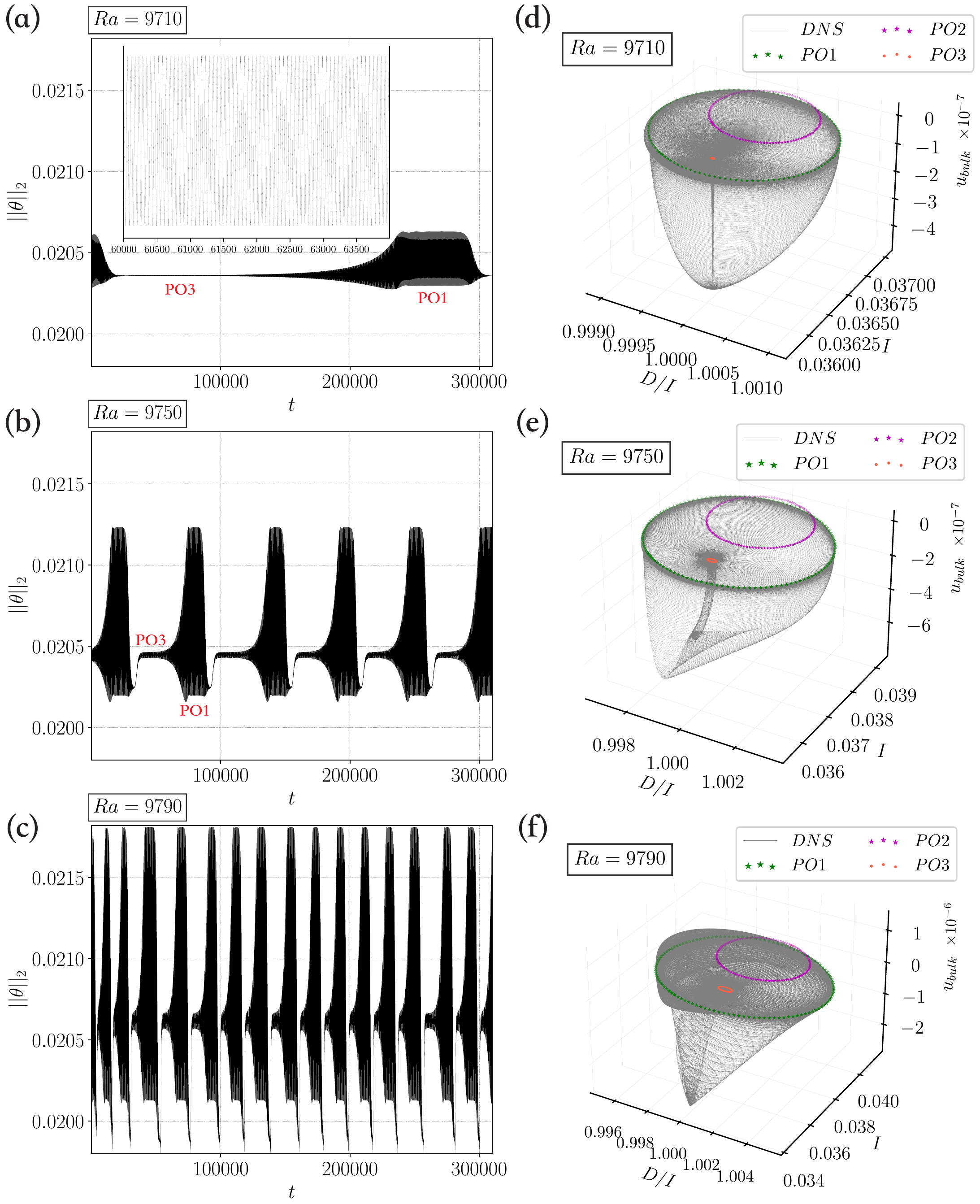}
    \captionsetup{font={footnotesize}}
    \captionsetup{width=13.5cm}
    \captionsetup{format=plain, justification=justified}
    \caption{\label{SDP_torus_series} (a--c) Time series for the long-time dynamics at $Ra=9710$, $9750$, and $9790$. The inset in (a) zooms in on the oscillatory behaviour, corresponding to PO3, between $60000\lesssim t \lesssim 64000$ with very small oscillation amplitude. (d--f) Phase portraits of the long DNS and three periodic orbits. The torus-like dynamics shows a heteroclinic cycle between two saddle orbits PO1 and PO3.}
\end{figure}

\par In addition to the simultaneously bifurcating branches that we investigated in \S \ref{SDP_simultaneous_bif}, \citet{Silber1991} also predicted a non-trivial attractor associated with regions of the normal-form coefficient space where all bifurcating branches are supercritical and unstable. This attractor is non-trivial because both the steady-state parent branch (FP1 in our system) and all bifurcating states are linearly unstable immediately after the Hopf bifurcation point. \citet{Silber1991} suggest that such bifurcations lead from a stable steady state directly to a structurally stable heteroclinic cycle, also called an invariant torus by \citet{Swift1988} who made the same prediction.

\par In order to confirm the theoretical predictions, we performed DNS initialised from random initial conditions and ran it for very long intervals ($\approx$$3\times 10^5$ time units). The DNS are conducted at three successively increasing Rayleigh numbers $Ra=9710$, $9750$, and $9790$, all above the critical value of Hopf bifurcation point ($Ra=9692.2$). Recall that the period of PO1, PO2, and PO3 are all around $T=250$ at these $Ra$ ranges (see figure \ref{SDP_TRTO_BD}b and table \ref{SDP_table_stability5states}). The time series of these simulations are shown in figures \ref{SDP_torus_series}(a--c). In each time series, two periodic behaviours are observed---one with small oscillation amplitude, corresponding to PO3, and another with large oscillation amplitude, corresponding to PO1---and there is a heteroclinic connection between these two periodic flows. When approaching the Hopf bifurcation point from above, the time spent near each of the two periodic orbits increases, and we expect that this time approaches infinity at the bifurcation point.

\par To visualise the complex dynamics and geometry of the non-trivial attractor, we project the flow fields onto $D$, $I$, and $u_{\text{bulk}}$ (defined in \S \ref{SDP_system_method}), in figures \ref{SDP_torus_series}(d--f). The simulation trajectory on the attractor is represented by continuous grey curves, and periodic orbits are illustrated by closed loops using coloured markers. These phase-space projections show torus-like structures where the trajectory spirals when leaving the saddle periodic orbits PO1 and PO3 along their unstable manifolds and when approaching them along their stable manifolds. For each case, the torus structure comprises two types of oscillations---a cone-like outer one with large amplitude and a chimney-like inner one with small amplitude. (The chimney structure is more difficult to identify visually at $Ra=9790$.) When leaving PO1 and approaching PO3, the PO1-type oscillations first spiral down (decrease of $u_{\text{bulk}}$) with decreasing oscillation amplitude, then rebound upwards (increase of $u_{\text{bulk}}$) towards PO3 along the chimney structure, with slightly increasing oscillation amplitude. Afterwards, when leaving PO3 and approaching PO1, the PO3-type oscillations spiral up (slight increase of $u_{\text{bulk}}$) with increasing oscillation amplitude.

\begin{table}
    \newcommand{\tabincell}[3]{\begin{tabular}{@{}#1@{}}#2\end{tabular}}
    \centering
    \begin{tabular}{cccccc}
    & $Ra=9710$ & \\
    State & $T$ & $N^u$ & Re$(\lambda_1)$ & $\sum_{i=1}^{N}$Re$(\lambda_i)$ & \\[0.2cm]
    PO1 & 231.242 & 2 & 0.000204 & 0.000205& \\
    PO2 & 232.198 & 4 & 0.000272 & 0.001087& \\
    PO3 & 234.958 & 2 & 0.000028 & 0.000034& \\
    TW1 & --& 4 & 0.000280 & 0.000688 & \\
    TW2 & --& 4 & 0.000347 & 0.001092& \\
    \end{tabular}
    \begin{tabular}{cccccc}
     $Ra=9750$& \\
     $T$ & $N^u$ & Re$(\lambda_1)$ & $\sum_{i=1}^{N}$Re$(\lambda_i)$ &  \\[0.2cm]
     232.861 &2 & 0.000639 & 0.000641& \\
     235.480 &4 & 0.000922 & 0.003688& \\
     243.844 &2 & 0.000219 & 0.000226& \\
     -- &4 & 0.001233 & 0.003285& \\
     -- &6 & 0.001179 & 0.004839& \\
    \end{tabular}
    \captionsetup{font={footnotesize}}
    \captionsetup{width=13.5cm}
    \captionsetup{format=plain, justification=justified}
    \caption{Properties of five simultaneously bifurcating states at $Ra=9710$ and $9750$. For each state, we list the period ($T$, for periodic orbits), the dimension of the unstable manifold ($N^u$), the Floquet exponent with the largest positive real part (Re$(\lambda_1)$), and the sum of the real parts of all unstable Floquet exponents ($\sum_{i=1}^{N}$Re$(\lambda_i)$).}
    \label{SDP_table_stability5states}
\end{table}

\par Note that PO2, TW1, and TW2 are not visited by the heteroclinic cycle. In order to relate this observation to the stability properties of the five states, we perform a linear stability analysis for all five states at $Ra=9710$ and $9750$; the results are summarised in table \ref{SDP_table_stability5states}. Indeed, at both Rayleigh numbers, PO1 and PO3 have fewer unstable eigendirections, a smaller leading unstable Floquet exponent, and a smaller sum of the positive real parts of the Floquet exponents than the other three states. Particularly, the leading Floquet exponent of PO3 is almost an order of magnitude smaller than other states at $Ra=9710$, and 3--5 times smaller at $Ra=9750$. These results reflect and are consistent with the previous observation (heteroclinic cycle between only PO1 and PO3), and explain why the heteroclinic cycle leaves PO3 much more slowly than leaving PO1; equivalently, the time spent near PO3 is longer than PO1, as already seen in figures \ref{SDP_torus_series}(a,b).

\subsection{Breaking of the heteroclinic cycle}
\begin{figure}
    \centering
    \includegraphics[width=\columnwidth]{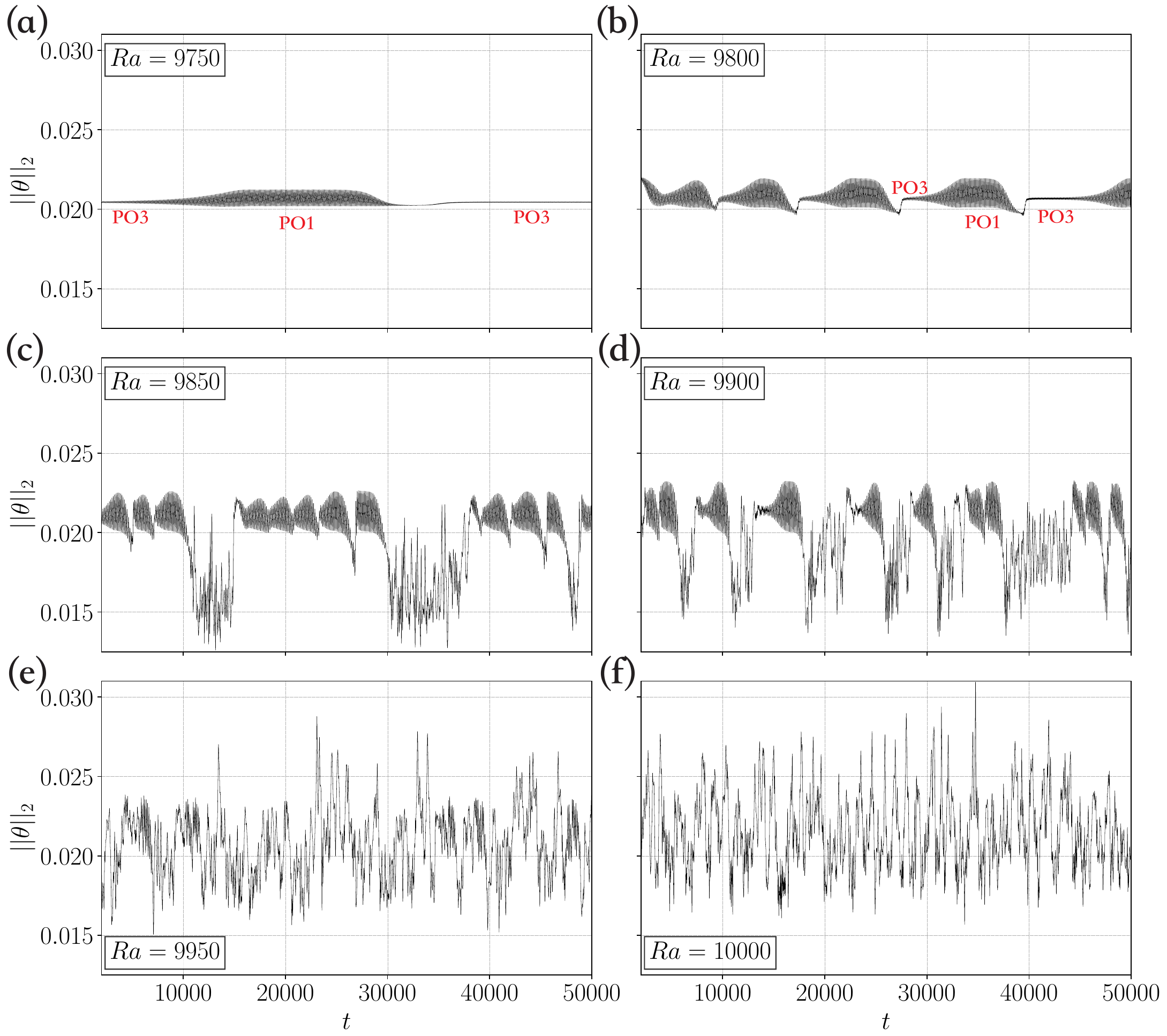}
    \captionsetup{font={footnotesize}}
    \captionsetup{width=13.5cm}
    \captionsetup{format=plain, justification=justified}
    \caption{\label{SDP_timeseries9750_10000} Time series of the long-time dynamics at $Ra=9750$, $9800$, $9850$, $9900$, $9950$, and $10000$. From low to high Rayleigh numbers, the quasi-periodic signal that corresponds to a heteroclinic cycle between PO1 and PO3, is gradually replaced by chaotic signals. Several temperature field snapshots at $Ra=9800$ and $Ra=10000$ are previously shown in figures \ref{SDP_large_domain}(e--f).}
\end{figure}

\par The heteroclinic cycle forms at the Hopf bifurcation point where FP1 loses stability, and we have seen in figure \ref{SDP_torus_series} that it persists until at least $Ra=9790$. However, this cycle should break at some point, so that more complex dynamics such as those shown in figure \ref{SDP_large_domain}(f) can take place. We perform DNS at higher Rayleigh numbers, up to $Ra=10000$, for an interval of $\approx$$5\times 10^4$ time units; the time series are shown in figure \ref{SDP_timeseries9750_10000}. Note that figure \ref{SDP_timeseries9750_10000}(a) is the same as figure \ref{SDP_torus_series}(b); the only difference is the scale of the $y$-axis.

\par At $Ra=9800$, the cycle still persists; the scenario is similar to the case of $Ra=9790$ in figure \ref{SDP_torus_series}(c). Two snapshots of the temperature field were previously shown in figure \ref{SDP_large_domain}(e). At $Ra=9850$, the system no longer exhibits an attracting heteroclinic cycle, and chaotic signals are observed for long time intervals at around $t\approx12000$ and $t\approx32000$. These chaotic episodes are associated with a significant decrease in the temperature norm $\| \theta \|_2$. Since PO1, PO3, and all other solutions discussed above consist of 12 (similar) rolls or small-amplitude oscillations of rolls, the decrease of the temperature norm during chaotic episodes suggests that the strength of rolls attenuates and/or the number of rolls decreases. We have subsequently verified (by checking the flow fields) that these chaotic signals correspond precisely to the onset of defects and stripes. (Determining the exact Rayleigh number between $Ra=9800$ and $9850$ for the onset lies outside of the scope of the present work.) At $Ra=9900$, the portion of the chaotic signals in the entire time series increases, yet (quasi-)periodic behaviour is still present. At $Ra=9950$, and further at $Ra=10000$, the flow transitions to a fully chaotic state; six snapshots at $Ra=10000$ were previously shown in figure \ref{SDP_large_domain}(f).

\par In \S \ref{SDP_TO1TO2}, we discussed the non-generic bifurcation structure resulting from the secondary state FP1. Even though PO1, PO2, PO3, TW1, and TW2 all capture local, small-amplitude modulations of convection rolls, it is unlikely that the analysis based on the linear instability of FP1 builds up the full complexity of the observed SDP pattern. The SDP pattern is observed with topological reorganization of the rolls in experiments \citep{Daniels2000} and large-domain simulations (figure \ref{SDP_large_domain}). As seen in figure \ref{SDP_timeseries9750_10000}, the heteroclinic connection breaks and more complex dynamics starts to play a role at around $Ra=9850$. These features cannot be explained entirely by the five invariant solutions discussed above. This suggests that there may exist other dynamically relevant solutions which can capture and explain other aspects of the SDP pattern. These solutions will be explored in the following sections.

\section{Global homoclinic bifurcation from ribbons} 
\label{SDP_orbit_SDP}
\begin{figure}
    \centering
    \includegraphics[width=\columnwidth]{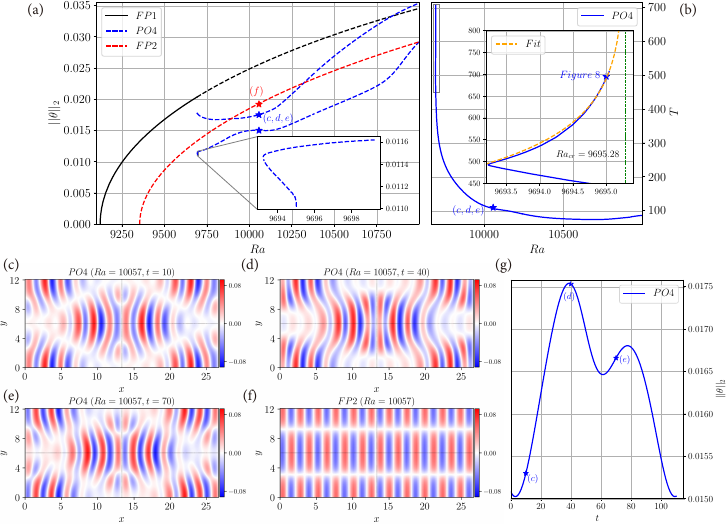}
    \captionsetup{font={footnotesize}}
    \captionsetup{width=13.5cm}
    \captionsetup{format=plain, justification=justified}
    \caption{\label{SDP_BD_snap_SDP} (a) Bifurcation diagram. Solid and dashed curves indicate stable and unstable states, respectively. The inset zooms in on the saddle--node bifurcation of PO4 at $Ra=9693.23$. (b) Periods of PO4. The inset zooms in on the range $9693 \lesssim Ra \lesssim 9695.5$, close to which a global homoclinic bifurcation occurs. The inset also shows the logarithmic fitting $T \approx -\frac{1}{|\lambda_4|}\ln(Ra_{cr}-Ra)+c_T$, with $\lambda_4 = 0.0103$, $Ra_{cr} = 9695.28$ and $c_T = 566$. Figure \ref{SDP_SDP_homo_phase_portrait} shows the analysis of PO4 on the upper branch at $Ra=9695$, marked on the solution curve. (c--f) Snapshots of the midplane temperature field of PO4 and FP2. The same colour bar is used in all plots. (g) Time series of PO4 at $Ra=10057$. In panels (a,b,g), the stars and labels indicate the locations of the visualisations of (c--f).}
\end{figure}

\par
\begin{figure}
    \centering
    \includegraphics[width=\columnwidth]{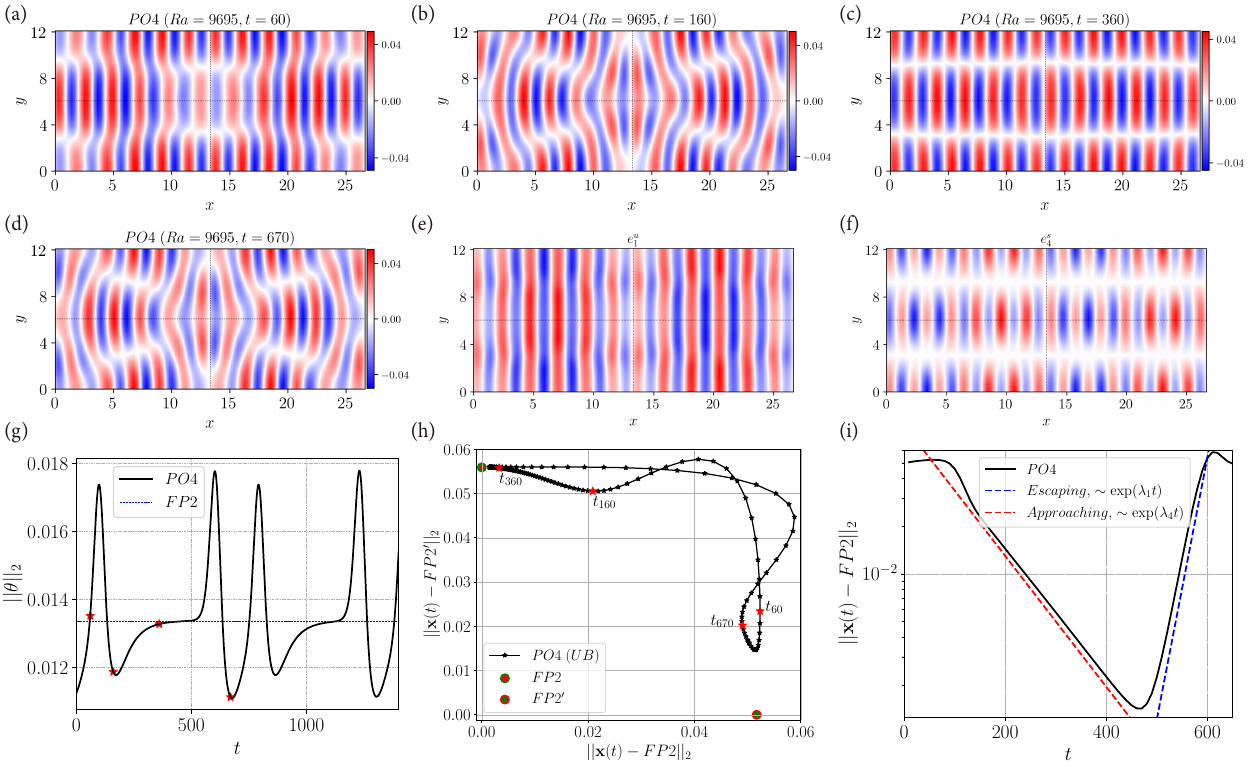}
    \captionsetup{font={footnotesize}}
    \captionsetup{width=13.5cm}
    \captionsetup{format=plain, justification=justified}
    \caption{\label{SDP_SDP_homo_phase_portrait} Orbit PO4 at $Ra = 9695$ with period $T=691.9$, on the upper branch. (a--d) Snapshots of the midplane temperature. Snapshots (b) and (d) show diamond-shaped amplitude modulations. Snapshots (a) and (c) are used as initial guesses in Newton's method to converge to FP2$^\prime$ and FP2. (e,f) Snapshots of the unstable and stable eigenmodes $e_{1}^u$ and $e_{4}^s$ of FP2, which are responsible for escaping from and approaching towards this equilibrium solution, respectively. (g) Time series from DNS. The four red stars indicate the moments where the snapshots (a--d) are taken. (h) Phase portrait of PO4, generated using the respective $L_2$-distances of PO4 from FP2 and FP2$^\prime$: $\|\boldsymbol{x}(t)-\text{FP}\|_2$. The points are clustered near two equilibria; FP2 is closely visited while FP2$^\prime$ is only partially visited. (i) $L_2$-distance between PO4 and FP2 as a function of time. The dynamics of PO4 is exponential for most of the cycle (black). The exponential approach to FP2 (red) and escape from it (blue) can be approximated using the eigenvalues $\lambda_{4}$ (slow approach rate) and $\lambda_{1}$ (fast escape rate), computed from the linearised dynamics around FP2.}
\end{figure}

 In this section, we discuss a periodic orbit PO4, shown in figures \ref{SDP_BD_snap_SDP}(c--e) at $Ra=10057$ and in figures \ref{SDP_SDP_homo_phase_portrait}(a--d) at $Ra=9695$. Orbit PO4 captures features observed in the large-domain simulations and experiments: the diamond-shaped amplitude modulations, rolls' defects, as well as the appearance and disappearance of diamond panes. Orbit PO4 has spatial symmetries $S_{\text{PO4}} \equiv \braket{\pi_y, \pi_{xz}, \tau (L_x/2, L_y/2)}$, and is linearly stable at $Ra=10057$ within this symmetry subspace. As a result, we can identify PO4 at this Rayleigh number using a DNS constrained to $S_{\text{PO4}}$. However, PO4 is linearly unstable throughout its range of existence in $Ra$ when perturbations in the full state space, breaking the symmetries $S_{\text{PO4}}$, are allowed.

\par As shown in figure \ref{SDP_BD_snap_SDP}(a), we have continued PO4 forward up to $Ra \approx 11000$. Continuing PO4 backwards, it undergoes a saddle--node bifurcation at $Ra = 9693.23$ (emphasised in the inset of figure \ref{SDP_BD_snap_SDP}a). This saddle--node bifurcation leads to the so-called upper branch because of its longer periods; see the inset of figure \ref{SDP_BD_snap_SDP}(b) where the period is visualised as a function of $Ra$. Within the subspace $S_{\text{PO4}}$, PO4 is stable for $9693.23 \le Ra \lesssim 10375$ (lower branch); the stability is lost after undergoing the saddle--node bifurcation and the instability at $Ra\approx10375$ is not discussed here. The upper branch then disappears in a global homoclinic bifurcation at $Ra_{cr} = 9695.28$.

\par Close to the global bifurcation point at $Ra = 9695$ on the upper branch, the dynamics of PO4 is relatively slow during a large portion of the orbit period, as illustrated by the plateau ($300\lesssim t \lesssim 500$) in the time series in figure \ref{SDP_SDP_homo_phase_portrait}(g). Two symmetry-related fixed points are identified by using the states at $t\approx60$ (shown in figure \ref{SDP_SDP_homo_phase_portrait}a) and at $t\approx360$ (shown in figure \ref{SDP_SDP_homo_phase_portrait}c) as initial guesses in Newton's method. These two fixed points capture the ribbons pattern and are called FP2 and FP2$^\prime \equiv \tau(L_x/2, 0)$FP2. Equilibrium FP2 at $Ra=10057$ is shown in figure \ref{SDP_BD_snap_SDP}(f). This equilibrium solution bifurcates from the unstable base flow at $Ra=9354.76$ (see figure \ref{SDP_BD_snap_SDP}a), remains unstable for its entire $Ra$-range of existence, and has the spatial symmetries $S_{\text{FP2}} = \braket{\pi_y, \pi_{xz}, \tau(L_x/11,0), \tau(L_x/22,L_y/2)}$.

\par To quantitatively characterise and visualise how FP2 and FP2$^\prime$ are visited by PO4 at $Ra = 9695$, we show a phase portrait in figure \ref{SDP_SDP_homo_phase_portrait}(h). This phase portrait depicts the distance between instantaneous flow fields along PO4 and the two fixed points: $\|\boldsymbol{x}(t)-\text{FP}\|_2$ where $\boldsymbol{x}(t)$ stands for the state vectors of PO4. We observe that many points, that are equally spaced in time, cluster near the two equilibria, and PO4 visits FP2 and FP2$^\prime$ with a minimum distance of the order of $10^{-3}$ and $10^{-2}$, respectively. Therefore, figures \ref{SDP_SDP_homo_phase_portrait}(g,h) together suggest that PO4 is close to a homoclinic orbit closely visiting FP2, and that FP2$^\prime$ does not play an essential role in the global bifurcation.

\par A homoclinic orbit approaches an equilibrium along one of its stable directions and escapes from it along one of its unstable directions. We compute the eigenvalues and eigenvectors of FP2 in $S_{\text{PO4}}$; the six leading eigenvalues are $[\lambda_1, \lambda_{2,3}, \lambda_4, \lambda_5, \lambda_6]= [0.0382, -0.0086 \pm 0.00276i, -0.0103, -0.0178, -0.0385]$. There is only one unstable eigendirection; we confirmed numerically that perturbing FP2 in this direction triggers a transition following a homoclinic connection that visits the neighbourhood of FP2$^\prime$ and finally returns to the vicinity of FP2. The unstable eigenmode $e_1^u$ associated with $\lambda_1$, shown in figure \ref{SDP_SDP_homo_phase_portrait}(e), contains 12 rolls; while FP2 has 11 rolls. Adding a small amount of the spatially detuned $e_1^u$ to FP2 causes phase interference: the modulation amplitude increases if the eigenmode is in phase with FP2 and decreases if out of phase. Such an observation is reminiscent of a sideband (also called modulational) instability giving rise to the observed large scale diamond-shaped amplitude modulation with defects in figures \ref{SDP_SDP_homo_phase_portrait}(b) and (d).

\par The direction along which PO4 approaches FP2 is $e_4^s$, associated with $\lambda_{4}$. This is confirmed by subtracting FP2 from the instantaneous flow fields of PO4 and comparing the result to the eigenmodes, as well as by the fact that the approaching phase does not contain any oscillations (not a complex conjugate pair such as $\lambda_{2,3}$). The exponential approaching and escaping dynamics of PO4 with respect to FP2 can be characterised by two eigenvalues $\lambda_{1,4}$ of FP2. This is illustrated in figure \ref{SDP_SDP_homo_phase_portrait}(i); the quantity $\|\boldsymbol{x}(t) - \text{FP2}\|_2$ decreases and increases exponentially at rates $\lambda_{4}$ and $\lambda_{1}$, respectively. Note that in this configuration, the eigendirection $e_4^s$ is not the leading stable (least negative) one, which is non-generic for a homoclinic orbit; we have not been able to give an explanation for this result.

\par As $\lambda_1 >|\lambda_4|$, the homoclinic orbit PO4 (upper branch) is unstable, as already stated above. The only way a stable periodic orbit (lower branch) can terminate in such a global homoclinic bifurcation is by first losing linear stability in a fold. This fold has a destabilising effect and has to exist prior to the global bifurcation. Along the backward continuation of PO4 from $Ra=10057$, its period increases monotonically. After undergoing the saddle--node bifurcation and close to the global bifurcation, the period diverges on the upper branch and reaches $T=706.01$ at $Ra =9695.0423$; see figure \ref{SDP_BD_snap_SDP}(b). The period of PO4 is dominated by the time for approaching FP2, as shown in figure \ref{SDP_SDP_homo_phase_portrait}(i). We have fitted the numerical data of the upper branch to the formula $T \approx -\frac{1}{|\lambda_4|}\ln(Ra_{cr}-Ra) +c_T$ (see \citet{Gaspard1990} and \citet{Reetz2020} for references of the formula), where $c_T$ is a fitting constant and $Ra_{cr} = 9695.28$ is determined from the fit, as shown in the inset of figure \ref{SDP_BD_snap_SDP}(b).

\section{Edge states between two attracting periodic orbits}
\label{SDP_edgestate}
\par Orbits PO1 (\S \ref{SDP_PO1TW1}) and PO4 (\S \ref{SDP_orbit_SDP}) have the same spatial symmetries $\braket{\pi_y, \pi_{xz}, \tau(L_x/2, L_y/2)}$, and are both linearly stable within this symmetry subspace for specific Rayleigh number ranges. Since there are two distinct attractors, there should exist a saddle state (edge state) separating their basins of attraction. In this section, we will track the edge states between PO1 and PO4 and analyse their bifurcation structures.

\subsection{Edge-state tracking at two Rayleigh numbers}
\begin{figure}
    \centering   
    \subfloat[$Ra = 10057$ ($\epsilon\approx0.1$).] 
    {
        \begin{minipage}[t]{\textwidth}
            \centering        
            \includegraphics[width=\columnwidth]{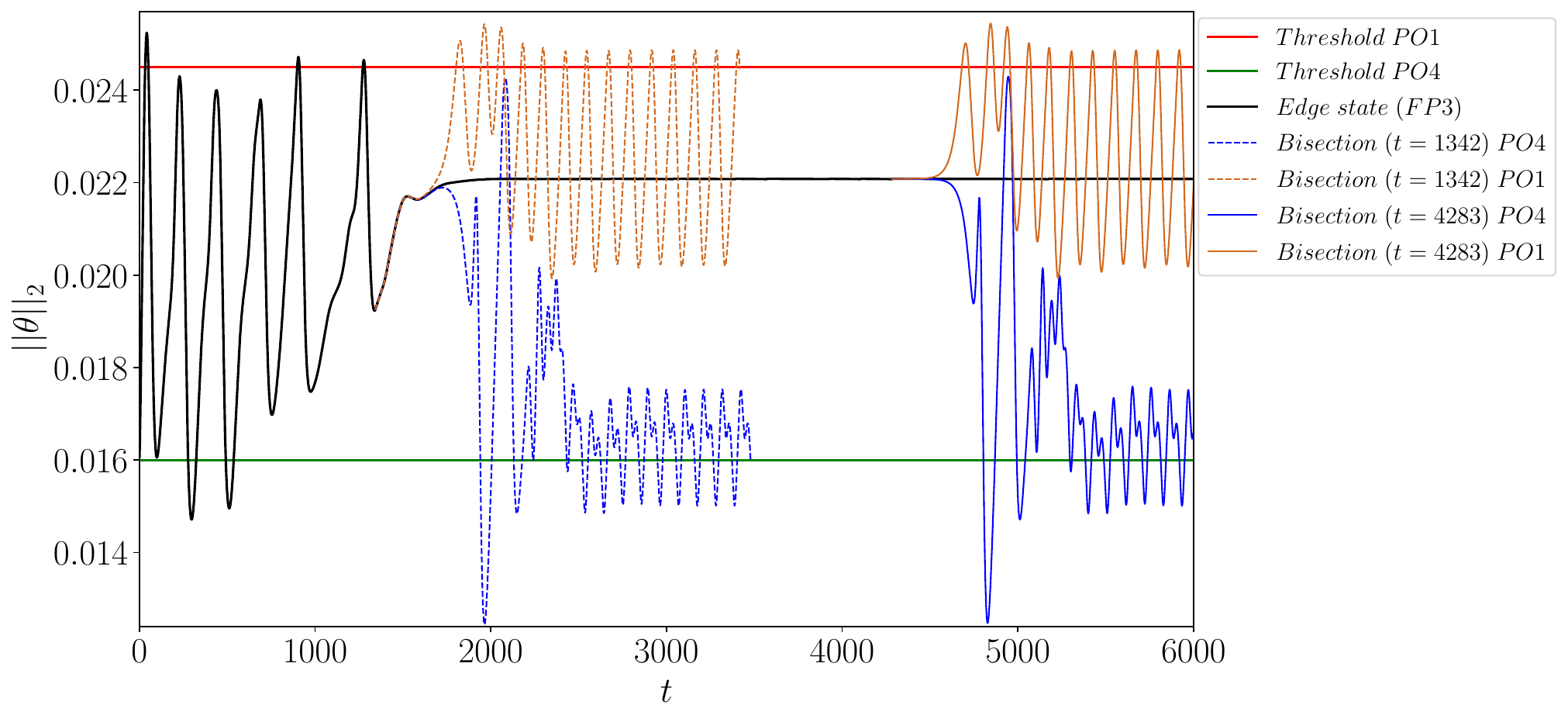}
            \captionsetup{font={footnotesize }}
        \end{minipage}
    }
    
    \subfloat[$Ra = 9800$ ($\epsilon\approx0.07$).] 
    {
        \begin{minipage}[t]{\textwidth}
            \centering    
            \includegraphics[width=\columnwidth]{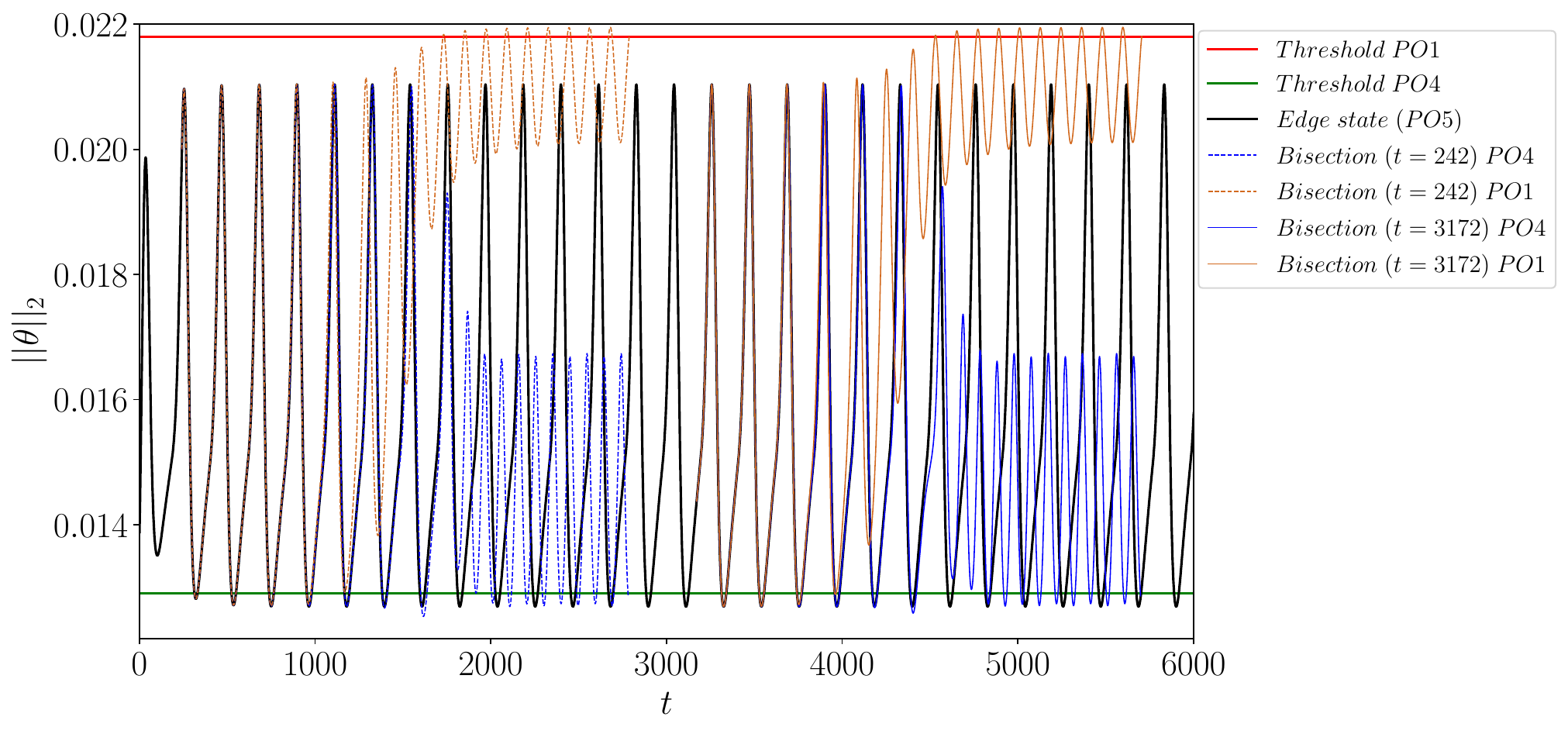} 
            \captionsetup{font={footnotesize }}
        \end{minipage}
    }
    \captionsetup{font={footnotesize}}
    \captionsetup{width=13.5cm}
    \captionsetup{format=plain, justification=justified}
    \caption{Time series from the edge tracking at two Rayleigh numbers $Ra=9800$ and $10057$. The case in (a) and (b) converges to an equilibrium (FP3) and a periodic orbit (PO5), respectively. The red and green horizontal lines represent the predefined upper and lower limits of $\| \theta \|_2$ characterising PO1 and PO4. The black curve corresponds to tracing the basin boundary until the respective edge state is reached after a sufficiently long time. Both figures include two bisection steps at two different instants.} 
    \label{SDP_edge_FP_PO}  
\end{figure}

\par In the literature, most of the edge trackings in shear flows have focused on turbulent--laminar attractors. The turbulent--laminar edge tracking can be traced back to the work of \citet{Skufca2006}, where the edge of chaos in parallel shear flows was discovered. Soon after, edge states were found in pipe flow \citep{Schneider2007b}, plane Couette flow \citep{Schneider2008b, Schneider2010a, Melnikov2014}, boundary layer flow \citep{Cherubini2011, Khapko2013, Kreilos2013a}, and plane Poiseuille flow \citep{Zammert2013}. In the context of the present work, in order to find the edge between two attracting periodic orbits, we use the temperature norm to distinguish between a PO4 attractor (for which $\| \theta \|_2$ fluctuates in a lower range) and a PO1 attractor (for which $\| \theta \|_2$ fluctuates in a higher range). According to the bifurcation diagram in figure \ref{SDP_TRTO_BD}(a) and figure \ref{SDP_BD_snap_SDP}(a), for $Ra \lesssim 10300$, the norms $\| \theta \|_2$ of PO1 and PO4 do not overlap and are clearly different.

\par To start the edge-tracking program, two initial conditions (flow fields)---one whose DNS trajectories converge to PO1 and the other to PO4---should be provided. The bisection method is adopted to find two extremely close states on either side of the edge. We then integrate these two states forward in time, with the edge always lying between them. After some time, the two resulting trajectories will separate in the edge’s unstable direction, and a refining bisection is required to determine a closer pair of states. The algorithm consists of a constant switching between the flow fields' bisection and forward time integration, to trace the stable manifold of the edge state---the basin boundary. In turbulent--laminar edge tracking, the time integration is continued until the cross-flow energy of the flow exceeds a predefined upper threshold or falls below a chosen lower threshold; then the current bisection is stopped, and a decision is made as to whether the state is turbulent or laminar. In our problem, however, a minimum integration of $2000$ time units is enforced before distinguishing between PO1 or PO4. The reason is to ignore the large variation in the temperature norm that occurs at the beginning of the time integration (initial transient) of each bisection (see figure \ref{SDP_edge_FP_PO}), and to let the dynamics settle onto the relevant orbit.

\begin{figure}
    \centering
    \includegraphics[width=0.45\columnwidth]{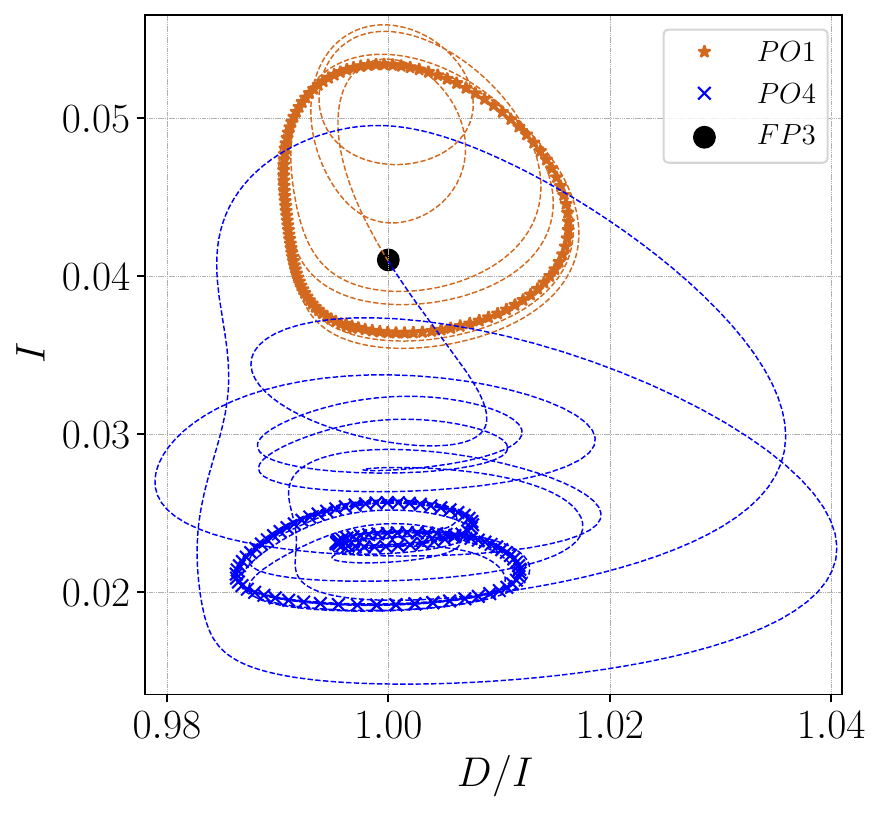}
    \captionsetup{font={footnotesize}}
    \captionsetup{width=13.5cm}
    \captionsetup{format=plain, justification=justified}
    \caption{\label{SDP_edge_TOSDP} Phase-space projection onto the thermal energy input ($I$) and the viscous dissipation over energy input ($D/I$), at $Ra = 10057$ ($\epsilon \approx 0.1$). The edge state equilibrium FP3 (black circle) is connected to PO1 (stars) and PO4 (crosses). Perturbing along different directions of the single unstable eigenvector of FP3 leads to either PO1 or PO4.}
\end{figure}

\par By running this algorithm, the program converges to a new equilibrium, that we will call FP3, at $Ra = 10057$, and a new periodic orbit, that we will call PO5, at $Ra = 9800$, as shown in figure \ref{SDP_edge_FP_PO}. Figures \ref{SDP_edge_FP_PO}(a) and (b) show the convergence to the edge state, along which several (only two are shown) bisections are performed to distinguish between PO1 and PO4. The upper and lower $\| \theta \|_2$ limits, indicated by red and green horizontal lines, are predefined thresholds used to determine whether the trajectory reaches PO1 or PO4. Both converged edge states are unstable with a single unstable eigenvector; the two directions along this vector connect the edge state to either PO1 or PO4, depending on the direction of the perturbation. This is illustrated in figure \ref{SDP_edge_TOSDP}, where the state space is visualised by a $2$D plane spanned by the rate of energy input ($I$) and dissipation ($D$); FP3 is the edge state at $Ra=10057$ with one unstable eigendirection, this direction connects FP3 to PO1 or PO4.

\subsection{Parametric continuations of edge states}
\par Once the edge states obtained by edge tracking are converged to invariant solutions by the Newton's method, parametric continuations in $Ra$ are performed to understand their bifurcation structures. The resulting bifurcation diagram is shown in figure \ref{SDP_BD_edge_states}(a).

\subsubsection{Bifurcation structure of FP3}
\begin{figure}
    \centering
    \includegraphics[width=\columnwidth]{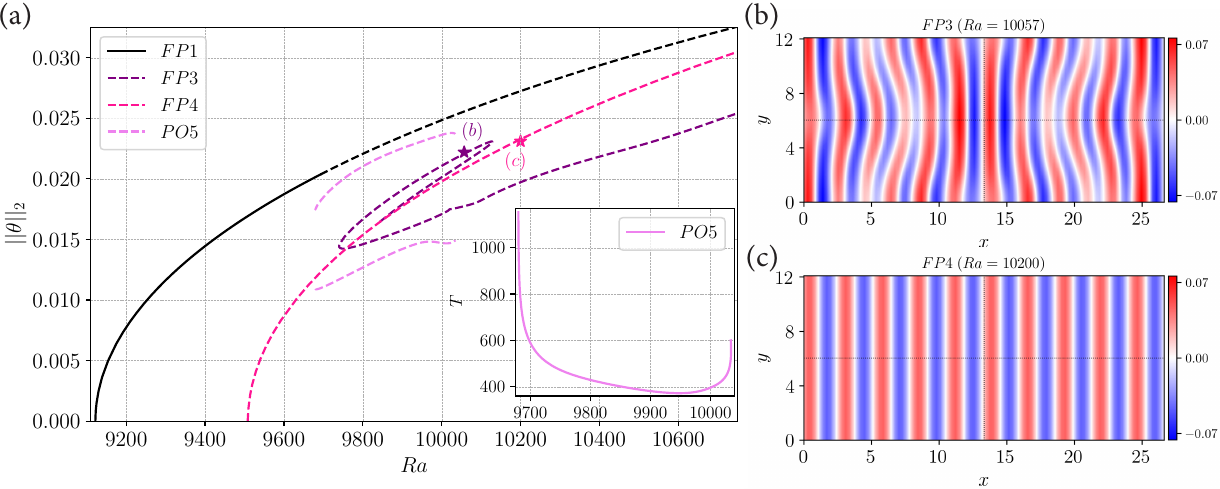}
    \captionsetup{font={footnotesize}}
    \captionsetup{width=13.5cm}
    \captionsetup{format=plain, justification=justified}
    \caption{\label{SDP_BD_edge_states} (a) Bifurcation diagram. Solid and dashed curves indicate stable and unstable states, respectively. The inset shows the periods of PO5 which diverge at its two ends. The stars and labels indicate the locations of the visualisations of (b--c). (b,c) Snapshots of the midplane temperature field of FP3 and FP4.}
\end{figure}

\par The edge state FP3, shown in \ref{SDP_BD_edge_states}(b), bifurcates from FP4 at $Ra=9843$ in a pitchfork bifurcation. Equilibrium FP4, shown in \ref{SDP_BD_edge_states}(c), in turn, bifurcates from the unstable base flow at $Ra=9508$. Equilibrium FP4 contains ten identical straight rolls; the wavelength of each roll is slightly longer than that of FP1. Equilibrium FP4 has the spatial symmetry $S_{\text{FP4}} \equiv \braket{\pi_{xz},\pi_y,\tau(L_x/10,\Delta y)}$; FP3 has the spatial symmetry $S_{\text{FP3}} \equiv \braket{\pi_{xz}, \pi_y, \tau(L_x/2, L_y/2)}$, the same symmetry group as PO1 and PO4. After bifurcating from FP4, FP3 undergoes two saddle--node bifurcations at $Ra=10128$ and then at $Ra = 9739.9$, and continues to exist beyond $Ra=11000$. Both FP3 and FP4 are linearly unstable for their entire existence range in $Ra$.

\par Equilibrium FP3 contains 12 rolls; the increase of number of rolls (from 10 to 12) along the FP3 branch with saddle--node bifurcations is reminiscent of the Eckhaus instability, e.g. the roll-creation and roll-splitting described in \citet{Zheng2024part1}. Equilibrium FP3 is an edge state between PO1 and PO4 only in the range $9954\lesssim Ra \lesssim 10111$, in the upper most branch in terms of $\| \theta \|_2$ projection; on all other parts of the branch, the perturbation always leads to PO4 or PO1.

\subsubsection{Bifurcation structure of PO5}
\begin{figure}
    \centering
    \includegraphics[width=\columnwidth]{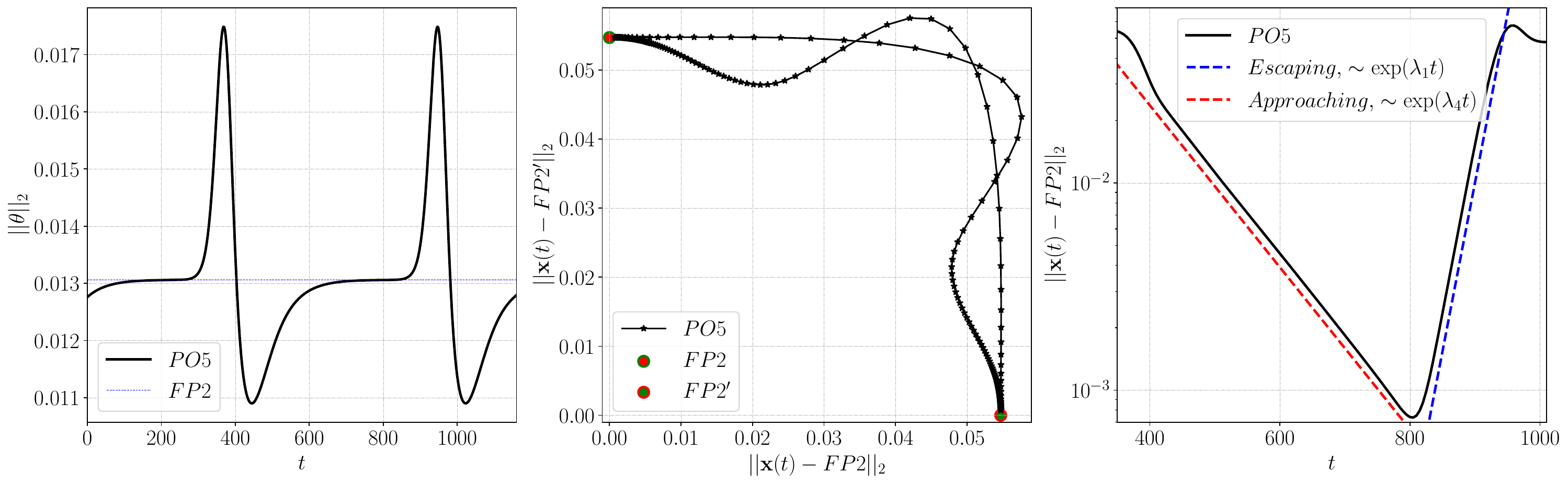}
    \captionsetup{font={footnotesize}}
    \captionsetup{width=13.5cm}
    \captionsetup{format=plain, justification=justified}
    \caption{\label{SDP_edge_state_po} Orbit PO5 at $Ra=9679.916$, close to the global bifurcation point. Left: time series from DNS. Middle: phase portrait with respect to FP2 and FP2$^\prime$. The points are clustered near two equilibria which are closely visited by PO5. Right: $L_2$-distance between PO5 and FP2. The dynamics of PO5 is exponential for most of the cycle (black line), the approach (red) and escape (blue) dynamics with respect to FP2 are shown and are governed by two eigenvalues of FP2.}
\end{figure}

\par The edge state PO5 does not show distinct features compared to PO4, thus we do not show its snapshots. It has the same spatial symmetries as PO4: $S_{\text{PO5}} \equiv \braket{\pi_{xz}, \pi_y, \tau(L_x/2, L_y/2)}$, and an additional spatio-temporal symmetry:
\begin{align}
    (u,v,w,\theta)(x,y,z,t+T/2) = (u,v,w,\theta)(x+L_x/2,y,z,t),
\end{align}
where $T$ is the period of PO5 at a given $Ra$. We continue PO5 backward and forward to $Ra=9679.916$ and $Ra=10034.285$, respectively. At its two ends, the period diverges as is shown in the inset of figure \ref{SDP_BD_edge_states}(a), suggesting two global bifurcations via which PO5 emerges or terminates. 

\par The time series and phase portrait of PO5 at $Ra=9679.916$ are shown in figure \ref{SDP_edge_state_po}; it can be seen that both FP2 and FP2$^\prime$ are closely visited by PO5. As is the case for PO4, the exponential approaching and escaping dynamics of PO5 with respect to FP2 are governed by two eigenvalues $\lambda_{1,4}$ of FP2. The associated eigenmodes $e_{1,4}$ are similar to those shown in figures \ref{SDP_SDP_homo_phase_portrait}(e,f). The minimum relative distance between PO5 and the two equilibria is of the order of $8\times 10^{-3}$. These pieces of evidence suggest that PO5 emerges from a heteroclinic cycle between two symmetrically-related versions of FP2. Interestingly, PO5 emerges when FP1 (secondary state) is still linearly stable. We do not discuss the global bifurcation close to $Ra=10034.285$ to avoid repetitions, as most features of the orbit and the analysis are the same as those discussed above.

\section{Summary and concluding remarks}
\label{SDP_conclusion}
\par The ILC system exhibits a profusion of flow regimes and self-organized coherent patterns, which makes this system a paradigm to study pattern formation. Like other shear flows, ILC exhibits a large-scale oblique pattern---switching diamond panes---which is reminiscent of the turbulent--laminar stripes in plane Couette flow. In the present study, by performing numerical simulations and continuations, we identify, within the framework of dynamical systems approach, multiple invariant states of the fully non-linear 3D Oberbeck–Boussinesq equations underlying the spatio-temporally complex SDP pattern, and construct their bifurcation diagram. These invariant states are found via time-dependent simulations (with possible simplifications by imposing symmetry constraints rendering solutions less unstable), recurrent flow analysis, and edge-state tracking. Generally, two types of bifurcation structures of these invariant solutions are identified and studied in detail: local (pitchfork, Hopf and saddle--node) bifurcations and global (homoclinic and heteroclinic) bifurcations.

\par Regarding local bifurcations, we investigate the linear instability of the twelve-roll branch FP1 and discover four new branches (including two periodic orbits and two travelling waves) that, together with the PO1 branch reported by \citet{Reetz2020}, bifurcate simultaneously from FP1. We show that the spatial symmetries of FP1 are responsible for the quadrupling of the eigenvalues at the bifurcation; the simultaneous creation of these five branches in supercritical Hopf bifurcations is understood as a consequence of breaking the spatial $[D_{4}]_{xz}$ and $[O(2)]_y$ symmetries. Above the bifurcation point, a heteroclinic cycle between two unstable periodic orbits is observed, and is interpreted as a consequence of these five unstable branches. Although phenomena resulting from the individual breaking of $D_{4}$ or $O(2)$ symmetries---simultaneous creation of two quantitatively different, not symmetrically-related branches---have been reported and analysed in the past, our scenario may be the first one reported in 3D hydrodynamic systems.

\par While the bifurcation structure at the Hopf bifurcation is already complicated, the resulting five states cannot fully explain the observed complex pattern. Both PO4 and PO5 are not part of the Hopf bifurcation but capture important features---diamond-shaped amplitude modulation and defect of rolls---of the SDP pattern. Orbits PO4 and PO5 are similar in terms of spatio-temporal features, but are different in terms of bifurcations; PO4 terminates by meeting FP2 in a homoclinic bifurcation, while PO5 terminates in a heteroclinic bifurcation by meeting two symmetrically-related copies of FP2. The modulational instability associated with FP2 or its unstable eigenmode is responsible for generating the large-scale diamond-shaped modulations in PO4 and PO5.

\par An edge tracking algorithm is employed as a tool not only for finding invariant solutions such as the edge states FP3 and PO5, but also for revealing the connection between the attractors---PO1 and PO4, in the same symmetry subspace. These two periodic orbits are not related by any local or global bifurcation, but are connected by an edge state separating their basins of attraction.

\par Besides exploring the complicated bifurcation scenarios of invariant solutions, from a pattern-forming point of view, the hope is that our solutions underlie the weakly turbulent dynamics and capture some of the coherent features of the flow. As discussed in \S \ref{SDP_large_simulation}, we start our investigation by performing large spatial domain simulations and then reduce it to a small domain---minimal flow unit supporting and containing the chaotic dynamics---for bifurcation analysis. Some features of the pattern or dynamics are lost in a small domain, but many important ones are preserved. Further imposing reflection symmetries in the small domain facilitated the identification of desired invariant solutions. Comparing snapshots of the converged invariant solutions in figures \ref{SDP_TRTO_BD}, \ref{SDP_BD_snap_SDP}, \ref{SDP_SDP_homo_phase_portrait} and \ref{SDP_BD_edge_states}, with those of the chaotic dynamics in figure \ref{SDP_large_domain}, it is tempting to suggest that our (symmetric) states capture many of the key features of the (non-symmetric) chaotic dynamics.

\par Among the 11 identified solutions, FP1 loses stability at $Ra=9692.2$ and all other ten solutions are linearly unstable in the $Ra$ range we consider. It has been conjectured that a turbulent flow can be seen as a trajectory ricocheting unstable invariant states in the phase space \citep{Kawahara2012}, and that a large number of unstable periodic orbits are believed to exist and to be embedded within the chaotic attractor of turbulence \citep{Chandler2013, Zheng2025part3}. The spatio-temporal dynamics of these invariant solutions may capture features of the trajectories that shadow them; this might also be true even when the solution itself has disappeared past a saddle--node bifurcation---the so-called ghost phenomenon, see \citet{Zheng2025ghost} for details. Describing and understanding weak turbulence by using non-linear dynamical systems concepts have been of great interest over the past decades \citep{Graham2021}, our results in the shear dominant region of transitional thermal convection emphasise its power in explaining the spatio-temporal chaos far from the threshold of linear instability.

\backsection[Acknowledgements]{We thank O. Ashtari for fruitful discussions and valuable comments on the manuscript.}
\backsection[Funding]{This work was supported by the European Research Council (ERC) under the European Union's Horizon 2020 research and innovation programme (grant no. 865677).}
\backsection[Declaration of interests]{The authors report no conflict of interest.}

\bibliographystyle{jfm}

\end{document}